\documentclass[11pt]{article}
\global\arraycolsep=1pt
\usepackage{amsmath,amssymb,graphicx}
\usepackage{hyperref}
\usepackage{multicol,color,longtable}
\definecolor{darkred}{rgb}{0.65,0.15,0}
\hypersetup{pdfborder={0 0 0},colorlinks=true,urlcolor=blue,citecolor=blue,linkcolor=darkred,linktocpage=true}

\usepackage{cite}

\usepackage{tikz}
\usetikzlibrary{arrows}
\usepackage{amsthm}
\usepackage{mathrsfs}
\usepackage[Symbol]{upgreek}
\usepackage{dsfont}
\usepackage[vcentermath]{youngtab}
\usepackage{textcomp}

\usepackage{latexsym}
\usepackage{graphicx}

\usepackage{calc}
\makeatletter
\newlength\@SizeOfCirc%
\newcommand{\CricArrowRight}[1]{%
    \setlength{\@SizeOfCirc}{\maxof{\widthof{#1}}{\heightof{#1}}}%
    \tikz [x=1.0ex,y=1.0ex,line width=.15ex, draw=black]%
        \draw [->,anchor=center]%
            node (0,0) {#1}%
            (0,1.2\@SizeOfCirc) arc (-240:85:1.2\@SizeOfCirc);%
}%
\makeatother

\newcommand{\sump}{\sideset{}{'}\sum}

\newcommand{\eprint}[1]{{\href{http://arxiv.org/abs/#1}{\texttt{[#1}]}}}
\newcommand{\eprintN}[1]{{\href{http://arxiv.org/abs/#1}{\texttt{#1 [hep-th]}}}}
\newcommand{\doi}[2]{{\hypersetup{urlcolor=darkred}\href{http://dx.doi.org/#2}{#1}\hypersetup{urlcolor=blue}}}
\newcommand{\eprintNM}[1]{{\href{http://arxiv.org/abs/#1}{\texttt{#1 [math]}}}}

\newcommand{\Tr}{\mathrm{Tr}\,}
\newcommand{\Tfrac}[2]{\scalebox{0.9}{$\frac{#1}{#2}$}}

\newcommand{\mf}[1]{{\mathfrak{#1}}}

\def\DJo{$\;$\kern-.4em \hbox{D\kern-.8em\raise.15ex\hbox{--}\kern.35em okovi\'c}}

\def\DEVII#1#2#3#4#5#6#7{{\tiny $ { \left[ \begin{array}{ccccccc}  & & \mathfrak{#2} \hspace{-0.7mm}&&&& \vspace{ -1.5mm} \\ \mathfrak{#1}\hspace{-0.7mm} &  \mathfrak{#3} \hspace{-0.7mm}& \mathfrak{#4} \hspace{-0.7mm} & \mathfrak{#5}\hspace{-0.7mm}&\mathfrak{#6}\hspace{-0.7mm}& #7 \hspace{-0.8mm} \end{array}\right] }$}}

\def\DEVIII#1#2#3#4#5#6#7#8{{\tiny $ { \left[ \begin{array}{ccccccc}  & & \mathfrak{#2} \hspace{-0.7mm}&&&& \vspace{ -1.5mm} \\ \mathfrak{#1}\hspace{-0.7mm} &  \mathfrak{#3} \hspace{-0.7mm}& \mathfrak{#4} \hspace{-0.7mm} & \mathfrak{#5}\hspace{-0.7mm}&\mathfrak{#6}\hspace{-0.7mm}&\mathfrak{#7}\hspace{-0.7mm}&\mathfrak{#8} \end{array}\right] }$}}

\newfont{\bbbold}{msbm10 scaled \magstep1}

\def\cE{{\cal E}}
\def\cF{{\cal F}}

\def\cH{{\cal H}}

\def\cL{{\cal L}}

\newfont{\goth}{eufm10 scaled \magstep1}

\def\p{\pi}

\def\be{\begin{equation}}\def\ee{\end{equation}}
\def\bea{\begin{eqnarray}}\def\eea{\end{eqnarray}}
\def\barr{\begin{array}}\def\earr{\end{array}}

\def\nn{\nonumber}
\def\bea{\begin{eqnarray}}
\def\eea{\end{eqnarray}}

\def\DEVI#1#2#3#4#5#6{{\tiny $ { \left[ \begin{array}{cccccc}  & & \mathfrak{#2} \hspace{-0.7mm}&&& \vspace{ -1.3mm} \\ \mathfrak{#1}\hspace{-0.7mm} &  \mathfrak{#3} \hspace{-0.7mm}& \mathfrak{#4} \hspace{-0.7mm} & \mathfrak{#5}\hspace{-0.7mm}& \mathfrak{#6} \end{array}\right] }$}}

\def\DSOX#1#2#3#4#5{{\tiny $ {   \biggl[ \begin{array}{ccc}  &&\mathfrak{#3}  \vspace{ -1.5mm} \\  \mathfrak{#1}\hspace{0.2mm}\mathfrak{#2}\hspace{-0.6mm} &\mathfrak{#4} \hspace{-0.9mm}&\vspace{-1.5mm}\\ && \mathfrak{#5}  \end{array}\biggr] }$}}

\def\det{{\rm det\,}}

\makeatletter

\@addtoreset{equation}{section} \makeatother

\newcommand{\scal}[1]{\bigl ({#1} \bigr )}

\newcommand{\CR}{\nonumber \\*}

\DeclareMathAlphabet{\mathpzc}{OT1}{pzc}{m}{it}

\newcommand{\gra}[2]{{\scriptscriptstyle (#1 , #2 )}}
\newcommand{\ord}[1]{{\scriptscriptstyle (#1)}}

\def\cL{{\cal L}}

\def\ie{{\it i.e.}\ }
\def\eg{{\it e.g.}\ }

\begin{document}

\thispagestyle{empty}

{\flushright {CPHT-RR094.122017}\\[15mm]}

\begin{center}
{\LARGE \bf Cancellation of divergences up to three loops\\[3mm] in exceptional field theory}\\[10mm]

\vspace{8mm}
\normalsize
{\large  Guillaume Bossard${}^{1,2}$ and Axel Kleinschmidt${}^{3,4}$}

\vspace{10mm}
${}^1${\it Centre de Physique Th\'eorique, Ecole Polytechnique, CNRS\\
Universit\'e Paris-Saclay 91128 Palaiseau cedex, France}
\vskip 1 em
{\it ${}^2$ Universit\`a di Roma Tor Vergata, Dipartimento di Fisica,\\
Via della Ricerca Scientifica, I-00133 Rome, Italy}
\vskip 1 em
${}^3${\it Max-Planck-Institut f\"{u}r Gravitationsphysik (Albert-Einstein-Institut)\\
Am M\"{u}hlenberg 1, DE-14476 Potsdam, Germany}
\vskip 1 em
${}^4${\it International Solvay Institutes\\
ULB-Campus Plaine CP231, BE-1050 Brussels, Belgium}

\vspace{20mm}

\hrule

\vspace{10mm}

\begin{tabular}{p{12cm}}
{\small
We consider the tetrahedral three-loop diagram in $E_d$ exceptional field theory evaluated as a scalar diagram for four external gravitons. At lowest order in momenta, this diagram contributes to the $\nabla^6 R^4$ term in the low-energy effective action for M-theory. We evaluate explicitly the sums over the discrete exceptional field theory loop momenta that become sums over 1/2-BPS states in the compact exceptional space. These sums can be rewritten as Eisenstein series that solve the homogeneous differential equations that supersymmetry implies for the $\nabla^6 R^4$ coupling. We also show how our results, even though sums over 1/2-BPS states, are consistent with expected 1/4-BPS contributions to the couplings.  
}
\end{tabular}
\vspace{7mm}
\hrule
\end{center}

\newpage
\setcounter{page}{1}

\setcounter{tocdepth}{1}
\tableofcontents

\vspace{5mm}
\hrule
\vspace{5mm}

\section{Introduction}

Determining the low-energy effective action of type II string theory compactified on a torus $T^{d-1}$ from $10$ to $D=11-d$ space-time dimensions has been an on-going research topic for many years~\cite{Green:1981yb,GrossWitten,Green:1997tv,Kiritsis:1997em,Obers:1999um,Green:1999pv,deWit:1999ir,Green:2005ba,Basu:2007ck,Basu:2007ru,Green:2008uj,Green:2010wi,Pioline:2010kb,Green:2010kv,Green:2011vz,Fleig:2012xa,Bossard:2014lra,Bossard:2014aea,Pioline:2015yea,Bossard:2015uga,Basu:2015dqa}. The low-energy effective action is interesting since one can hope to understand better how string theory improves the ultraviolet behaviour of point particle theories like gravity and supergravity through infinite towers of massive particles that restore unitarity at high energy.  Taking various limits of the exact couplings also provides important information on non-perturbative objects in string theory, like D-branes, membranes or black holes. They include in particular helicity supertraces or partition functions associated to solitons or instantons, respectively.  The effective action includes, besides the standard two-derivative action, an infinite set of higher-derivative corrections, {\it e.g}. of the form $\nabla^{2k} R^4$ in the case of four-graviton scattering. These couplings were originally obtained from the computation of the perturbative string scattering amplitudes of states belonging to the massless graviton supermultiplet, and their low-energy expansion in  $\alpha'=\ell_{\text{s}}^2$~\cite{GrossWitten,Green:1999pv,Green:2008uj}. With the (conjectural) discovery of non-perturbative U-duality $E_{d}(\mathds{Z})$ for the maximally supersymmetric compactifications on tori $T^{d-1}$~\cite{Hull:1994ys}, non-perturbative contributions to the higher-derivative corrections could be determined and are often related to automorphic forms. Together with an analysis of supersymmetry constraints~\cite{Green:1998by,Bossard:2014lra,Bossard:2014aea,Bossard:2015uga}, one can sometimes prove uniqueness of the perturbative and non-perturbative contributions to certain higher-derivative corrections~\cite{Pioline:1998mn}.

\begin{figure}
\centering
\begin{picture}(200,50)
\thicklines
\multiput(10,10)(30,0){4}{\circle*{10}}
\multiput(97,10)(10,0){5}{\line(1,0){5}}
\put(10,10){\line(1,0){90}}
\put(150,10){\line(1,0){30}}
\put(150,10){\circle*{10}}
\put(180,10){\circle*{10}}
\put(70,40){\circle*{10}}
\put(70,10){\line(0,1){30}}
\put(7,-5){$1$}
\put(57,38){$2$}
\put(37,-5){$3$}
\put(67,-5){$4$}
\put(97,-5){$5$}
\put(142,-5){$d\!-\!1$}
\put(177,-5){$d$}
\end{picture}
\caption{\label{fig:dynk}\small Dynkin diagram of $E_{d}$.}
\end{figure}
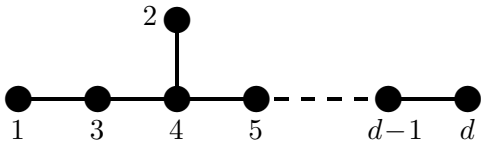

Using the above methods together with consistency relations coming from various perturbative and decompactification limits, it has been possible to pin down the $E_{d}(\mathds{Z})$-dependence of the correction terms $R^4$ and $\nabla^4R^4$ on the moduli $\Phi$ parametrising the symmetric space $E_{d}(\mathds{R})/K(E_{d})$, where $E_d(\mathds{R})$ is the split real form $E_{d(d)}$ and $K(E_d)$ its maximal compact subgroup. The Dynkin diagram of $E_{d(d)}$ is shown in Figure~\ref{fig:dynk}. One has to distinguish the cases $d\leq 7$ where the U-duality symmetry $E_{d}(\mathds{Z})$ follows naturally from charge quantisation, and the lattice of charges support the spectrum of BPS particles, from the cases $d=8,9$ for which there is no such interpretation. In these cases one has nonetheless conjectured U-dualities following the same pattern~\cite{Hull:1994ys,Mizoguchi:1999fu} and one can define a field theory in $D=3$ and $D=2$ dimensions, respectively, with a low-energy two-derivative action that exhibits $E_{d}(\mathds{R})$ symmetry. By $E_9$ we denote the affine Kac--Moody extension of $E_8$. For the indefinite hyperbolic and Lorentzian Kac--Moody symmetries $E_{10}$ and $E_{11}$, the dynamical theory is less clear and the definition and separation of massless amplitudes is ill-defined, but one can still formally use automorphic forms on them that are a book-keeping device in that they decompactify correctly to the cases $E_{d(d)}$ with $d\leq 9$~\cite{Fleig:2012xa}.
In general,  these corrections are given by certain Eisenstein series on the groups $E_{d(d)}$. Speaking in the language of automorphic representations, the $\tfrac12$-BPS correction $R^4$ belongs to the minimal unitary representation of $E_{d}(\mathds{R})$ while the $\tfrac14$-BPS correction $\nabla^4R^4$ belongs to a next-to-minimal unitary representation~\cite{Pioline:2010kb,Green:2011vz} (that is unique for $d\ge 7$). The next case $\nabla^6R^4$ has also attracted attention in the last years and has been treated using different methods in~\cite{Green:2005ba,Basu:2007ck,Green:2014yxa,Basu:2014hsa,Bossard:2015uga,Pioline:2015yea}. The corresponding function multiplying the $\nabla^6R^4$ term in the effective action, often denoted $\mathcal{E}_{\gra{0}{1}}$,  is generally the sum of two functions that correspond to two distinct  $\tfrac18$-BPS supersymmetry invariants \cite{Bossard:2015uga}. One is an Eisenstein series attached to a next-to-minimal representation for $d\le 6$ (and next-to-next-to-minimal for $d=7$), while the other satisfies an inhomogeneous differential equations with sources quadratic in $\mathcal{E}_{\gra{0}{0}}$ \cite{Green:2005ba}. Consequently, the latter is not an automorphic form and cannot be attached to an automorphic representation in the standard sense. Despite this, one can prove that its Fourier coefficients and the differential equations it satisfies are naturally associated to a nilpotent orbit \cite{Bossard:2015oxa}.

The coefficient functions $\mathcal{E}_{\gra{0}{1}}$ multiplying the $R^4$ term and $\mathcal{E}_{\gra{1}{0}}$ multiplying the $\nabla^4 R^4$ term for compactifications on $T^{d-1}$ were determined indirectly using consistency arguments. A direct calculation was undertaken recently in~\cite{Bossard:2015foa} and based on the framework of exceptional field theory. Exceptional field theory~\cite{Koepsell:2000xg,Hull:2007zu,Pacheco:2008ps,Berman:2010is,Berman:2011cg,Cederwall:2013naa,Hohm:2013pua,Hohm:2013vpa,Hohm:2013uia,Aldazabal:2013via,Godazgar:2014nqa,Hohm:2014fxa} in $D$ space-time dimensions uses an extended `internal' space whose coordinates $Y^M$ transform in a representation $\overline{\bf R}_{\alpha_d}$ of the symmetry group $E_{d}(\mathds{R})$ where $d=11-D$. In order to eliminate extra degrees of freedom compared to supergravity, any field of the theory (and product of fields) is required to satisfy a section constraint that transforms in a different representation $\overline{\bf R}_{\alpha_1}$ of $E_{d}(\mathds{R})$. These representations are tabulated for the various values of $0<d\leq 8$ in Table~\ref{tab:reps}. More precisely, one demands
\begin{align}
\frac{\partial}{\partial Y^M} A(x,Y) \frac{\partial}{\partial Y^N} B(x,Y) \bigg|_{\,\overline{\bf R}_{\alpha_1}} = 0
\end{align}
for any two fields $A(x,Y)$ and $B(x,Y)$, where $x^\mu$ are the standard $D$-dimensional coordinates and $Y^M$ the extended coordinates. The representation $\overline{\bf R}_{\alpha_1}$ is contained in the tensor product of two representations ${\bf R}_{\alpha_d}$ and can be interpreted as a $\tfrac12$-BPS constraint. When the theory is defined on $D$-dimensional Minkowski space times the exceptional torus ({\it i.e.} in a background independent of the $Y$ coordinates), the Fourier modes of momentum $\Gamma$ satisfying the $\tfrac12$-BPS constraint can be interpreted as massive $\tfrac12$-BPS supermultiplets of states. The exceptional field theory Lagrangian permits to describe the three-point interactions of these multiplets, and their coupling to the massless supermultiplet.

\begin{table}
\centering
\begin{tabular}{c|c||c|c}
Space-time dimension& Hidden symmetry & coordinates $Y^M$ & Section constraint\\
$D=11-d$ & $E_{d}(\mathds{R})$ & $\overline{\bf R}_{\alpha_d}$ & $\overline{\bf R}_{\alpha_1}$\\[2mm]\hline
$9$ & $GL(2,\mathds{R})$ & ${\bf 1}^\ord{-4}\oplus {\bf 2}^\ord{3}$ & ${\bf 2}^\ord{-1}$\\
$8$ & $SL(2,\mathds{R})\times SL(3,\mathds{R})$ & $({\bf 2}, {\bf 3})$ & $({\bf 1},\overline{\bf 3})$\\
$7$ & $SL(5,\mathds{R})$ & ${\bf 10}$ & ${\bf \overline{5}}$\\[2mm]
$6$ & $SO(5,5,\mathds{R})$ & ${\bf 16}$ & ${\bf 10}$\\[2mm]
$5$ & $E_{6}(\mathds{R})$ & ${\bf 27}$ & ${\bf \overline{27}}$\\[2mm]
$4$ & $E_{7}(\mathds{R})$ & ${\bf 56}$ & ${\bf 133}$\\[2mm]
$3$ &$E_{8}(\mathds{R})$ & ${\bf 248}$ & ${\bf 3875}$\\[2mm]
\end{tabular}
\caption{\label{tab:reps}\small Coordinate representation $\overline{\bf R}_{\alpha_d}$ and strong section constraint representation $\overline{\bf R}_{\alpha_1}$ for hidden symmetry groups $E_{d(d)}$ in dimension $D=11-d$ for $2\leq d \leq 8$.}
\end{table}

Using this formalism and explicit one- and two-loop calculations in exceptional field theory, together with a reduction to scalar diagrams as in~\cite{Bern:2007hh,Bern:2008pv}, we recovered from a direct calculation the $R^4$ and $\nabla^4R^4$ correction functions in~\cite{Bossard:2015foa}, confirming the previous indirect results. We also obtained a form of the $\nabla^6 R^4$ correction function consistent with its differential properties described above. Nonetheless, the consistency of our result required to neglect the one-loop contribution to the $\nabla^4R^4$ correction to avoid divergences and the doubling of the coefficient. Moreover, it is expected that the latter does not get contributions from higher loops but it is known that $\nabla^6 R^4$  is corrected at three-loop. The calculations in~\cite{Bossard:2015foa} can be seen as a U-duality completion of supergravity loop calculations carried out in~\cite{Green:1997as,Green:1999pu} by including full multiplets of $\tfrac12$-BPS states~\cite{deWit:1999ir}.

In the present paper, we extend this analysis to the three-loop contribution to the $\mathcal{E}_{\gra{0}{1}}$ coupling in exceptional field theory. As is known from~\cite{Bern:2007hh,Bern:2008pv}, there are several topologies of scalar diagrams that arise at three-loop order in maximal supergravity. Not all of them are amenable to the exceptional field theory techniques developed in~\cite{Bossard:2015foa}. However, only one of them is relevant for the $\nabla^6 R^4$ correction and it is treatable in perturbative exceptional field theory. The skeleton graph in this case has tetrahedral structure~\cite{Basu:2014hsa}.

By a careful analysis of the solutions of the section constraint and exploiting the symmetries of the tetrahedron, we shall derive automorphic functions that solve the relevant differential equations and we shall also see how our calculation exhibits a cancellation of divergences in the various dimensions, with a dependence in a renormalisation scale consistent with the known and expected ultraviolet divergences in supergravity.

We shall also discuss in detail aspects of the regularisation of the exceptional field theory amplitudes, expanding on our proposal in~\cite{Bossard:2015foa}. As mentioned above, the one-loop exceptional field theory contribution to the $\nabla^4R^4$ coupling  must be renormalised to zero to give the correct finite result.  The cancelling contribution was argued in~\cite{Bossard:2015foa} to come from the contributions of $\tfrac14$-BPS states that are neglected in exceptional field theory. In Section~\ref{sec:14BPS}, we shall argue that one can obtain these $\frac14$-BPS contributions by U-duality covariantisation of the $\nabla^4R^4$ coupling obtained from perturbative string theory at one-loop. We shall exhibit a formal cancellation of the $\tfrac12$-BPS states (coming from exceptional field theory) and the $\frac14$-BPS states contributions (from string theory), confirming the validity of the picture in~\cite{Bossard:2015foa}. We also extend these arguments and discuss more generally the systematics of BPS corrections up to $\nabla^6R^4$ in Section~\ref{sec:system} where we also discuss non-renormalisation properties of BPS solitons and instantons. This will allow us to exhibit that our framework provides a consistent approach to determining the low-energy behaviour of the four-graviton scattering process up to order $\nabla^6R^4$.

\section{The tetrahedral diagram and its symmetries}

Up to two loops, all the Feynman diagrams contributing to the four-graviton scattering amplitude involves internal momenta that satisfy the strong section constraint $\Gamma_i \times \Gamma_j = 0$. Here, $\Gamma_i$ for $i=1,2$ are the discrete charges of the supermultiplet circulating in the loops. The discrete charges are in the lattice $\mathds{Z}^{d(\alpha_d)}$ in the $E_{d}(\mathds{R})$ representation ${\bf R}_{\alpha_d}$ of dimension $d(\alpha_d)$ shown in Table~\ref{tab:reps}. As explained in \cite{Bossard:2015foa}, each contribution is then necessarily in the U-duality orbit of a supergravity amplitude in two more dimensions on $\mathds{R}^D\times T^2$. It then follows that the reduction of the amplitudes derived in supergravity in~\cite{Bern:2007hh,Bern:2008pv} applies, and the exceptional field theory amplitude reduces to the U-duality covariantisation ({\it i.e.} Poincar\'e sum over U-duality orbits) of the supergravity amplitude. At three loops this is no longer the case in general, and for example for the ladder diagram shown on the right of Figure~\ref{fig:skeletons}, the momenta do not necessarily satisfy the strong section constraint. Moreover, the amplitude includes then a priori four-point vertices between four massive states of charges satisfying the strong section constraint, and the non-associativity of the convolution product subjected to the section constraint implies that one cannot neglect the 1/4 BPS states multiplets in this computation. Nonetheless, we shall argue that the $\nabla^6 R^4$ coupling is still determined  by the tetrahedral diagram contribution only in exceptional field theory.

 \def\xshift{5}
  \def\xmin{1}
 \def\ymin{-2}
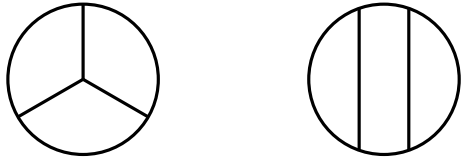
\begin{figure}[t!]
\begin{center}
 \begin{tikzpicture}
  \draw[-,draw=black,very thick] (\xmin-2,\ymin) circle (1cm);
    \draw[-,draw=black,very thick](\xmin-2,\ymin) -- (\xmin-2,\ymin+1);
        \draw[-,draw=black,very thick](\xmin-2,\ymin) -- (\xmin-2+0.866,\ymin-0.5);
                \draw[-,draw=black,very thick](\xmin-2,\ymin) -- (\xmin-2-0.866,\ymin-0.5);
  \draw[-,draw=black,very thick] (\xmin+2,\ymin) circle (1cm);
        \draw[-,draw=black,very thick](\xmin+2+0.33,\ymin-0.94) -- (\xmin+2+0.33,\ymin+0.94);
        \draw[-,draw=black,very thick](\xmin+2-0.33,\ymin-0.94) -- (\xmin+2-0.33,\ymin+0.94);
      \end{tikzpicture}
\end{center}
\caption{\small The skeletons graphs at three-loop order. The tetrahedral graph on the left starts contributing at order $\nabla^6 R^4$ in the derivative expansion whereas the ladder diagram on the right has its first contribution at order $\nabla^8R^4$.}
\label{fig:skeletons}
\end{figure}

There are two skeleton diagrams that arise at three-loop order, the tetrahedral graph (a.k.a. Mercedes diagram) and the ladder diagram depicted in Figure~\ref{fig:skeletons}. The ladder diagram does not give any contribution to order $\nabla^6 R^4$ in supergravity \cite{Bern:2008pv}. When all internal momenta satisfy the strong section constraint $\Gamma_i \times \Gamma_j = 0$, one can always use an element of the U-duality group to consider an equivalent representative in eleven-dimensional (or type IIB) supergravity, such that the same reduction of the diagrams computed in \cite{Bern:2008pv} also applies in exceptional field theory. At one loop and at two loops, the fact that the three-point vertices satisfy the section constraint implies  that all momenta $\Gamma_i$  have to satisfy the strong section constraint consistently in a pairwise manner. But at  three loops one can have diagrams (like the ladder diagram) for which among the three momenta satisfying $\Gamma_i \times \Gamma_i=0$, two of them fail to satisfy the strong section constraint, say $\Gamma_1 \times \Gamma_2 \ne 0$ while $\Gamma_1\times \Gamma_3 = \Gamma_2 \times \Gamma_3=0$, see also~\cite{Bossard:2015foa}. Such a configuration is not $E_{d}(\mathds{Z})$ equivalent to a configuration of momenta in supergravity and one cannot rely on \cite{Bern:2008pv} to deduce that they could only contribute to higher order derivative couplings. Nonetheless, these contributions with two momenta failing to satisfy the section constraint can be moved to a frame where one can see them as momenta and winding of perturbative strings on the torus. In this case indeed one can always find an element of $E_{d}(\mathds{Z})$ to rotate the element $\Gamma_1 \times \Gamma_2$ in ${\bf R}_{\alpha_1}$ to the highest weight representative. In the string perturbative parabolic decomposition $P_1$ of $E_{d}(\mathds{R})$\footnote{With $V$ we denote the vector representation of $\mf{so}(d-1,d-1)$ while $S_{\pm}$ denote the chiral spinors with the convention that $S_-$ has its non-zero highest weight label on the node attached to node $1$ when embedded in $\mf{e}_d$. For even $d$ one has $\overline{S}_- \cong S_-$ while for odd $d$ the isomorphism is $\overline{S}_- \cong S_+$.}
\bea
\hspace{-2mm}  \mathfrak{e}_{d(d)} &\cong& ( \Lambda^{d-7} V)^\ord{-2} \oplus {S}_-^\ord{-1} \oplus ( \mathfrak{gl}_1 \oplus \mathfrak{so}(d-1,d-1))^\ord{0} \oplus \overline{S}_-^\ord{1} \oplus ( \Lambda^{d-7} V)^\ord{2} \;, \CR
R(\Lambda_d) &\cong& \dots \oplus ( \Lambda^{d-8} V \oplus \Lambda^{d-6} V)^\ord{2\frac{d-8}{9-d}} \oplus S_+^\ord{\frac{d-7}{9-d}} \oplus V^\ord{\frac{2}{9-d}} \; , \CR
R(\Lambda_1) &\cong& \dots \oplus ( \Lambda^{d-7} V \otimes S_-)^\ord{\frac{3d-23}{9-d}} \oplus ( \Lambda^{d-7} V \oplus \Lambda^{d-5} V)^\ord{\frac{2d-14}{9-d}}  \oplus S_-^\ord{\frac{d-5}{9-d}} \oplus {\bf 1}^\ord{\frac{4}{9-d}} \ ,\hspace{2mm}  
\eea 
the only solutions $\Gamma_i$ to the section constraints compatible with the property that $\Gamma_1 \times \Gamma_2 \in {\bf 1}^\ord{\frac{4}{9-d}}$ are such that all $\Gamma_i \in V^\ord{\frac{2}{9-d}}$. One can always choose the representative such that $\Gamma_1 +\Gamma_2 \in V^\ord{\frac{2}{9-d}} $, and the constraint that $\Gamma_3 \times (\Gamma_1 +\Gamma_2) = 0$ and $(\Gamma_1-\Gamma_2) \times (\Gamma_1 +\Gamma_2) = 0$ impose that they both belong to $V^\ord{\frac{2}{9-d}} $ as well.  

The low-energy expansion of the three-loop 4-graviton scattering amplitude computed in \cite{Gomez:2013sla} can be extended straightforwardly to toroidal compactifications \cite{Pioline:2015yea}. The resulting string theory contribution to the $\nabla^6 R^4$ coupling is simply the integral over the genus-3 moduli space $Sp(6,\mathds{Z})\backslash Sp(6,\mathds{R})/ U(3)$ of the $SO(d-1,d-1)$ genus-3 Narain partition function. It follows that the only string states contributing to $\nabla^6 R^4$ at 3-loop are level matched, and so satisfy the strong section constraint. One can check in particular that this contribution is the $SO(d-1,d-1,\mathds{Z})$ covariantisation of the ten-dimensional supergravity amplitude as written in \cite{Basu:2014hsa}. We conclude therefore that there are no contributions to the $\nabla^6 R^4$ coupling that fail to satisfy the strong section constraint at 3-loops. In the following we shall therefore consider that the unique contribution comes from the tetrahedral diagram.

Here we assume therefore that the exceptional field theory integrand is identical to the one deriving from supergravity. The 3-loop supergravity amplitude was evaluated in~\cite{Bern:2007hh,Bern:2008pv} and it was shown in~\cite{Basu:2014hsa} that one can use several integrations by part to simplify the three-loop tetrahedral diagram integrand to a manifestly $SL(3,\mathds{Z})$ invariant integrand in nine dimensions.\footnote{Note that the individual non-amputated diagrams underlying the tetrahedral skeleton are not just of scalar $\phi^3$ type since they have non-trivial momentum dependence in the numerators.} The dual graph of the tetrahedral skeleton is the regular tetrahedron, making obvious that the symmetric group $S_4$ is a symmetry of the configuration and we will see below that the amplitude can be written in a way that is manifestly symmetric under this group \cite{Basu:2014hsa}. Extrapolating this result to exceptional field theory, one obtains the contribution to the effective action  at order $\nabla^6R^4$ 
 \begin{align}
\label{eq:a1}
\mathcal{E}_{\gra{0}{1}}^\ord{\textrm{3-loop}} = \frac56 \sum_{\substack{\Gamma_i\in \mathds{Z}^{ 3d(\alpha_d)}\\ \Gamma_i\times \Gamma_j =0}} \int_{S^+} \frac{d^6\Omega}{\left(\det \Omega\right)^{2-\frac{d-5}{2}}} \exp_\mu\Big( -\pi \Omega^{ij} g(\Gamma_i,\Gamma_j)\Big)\,.
\end{align}
Here, each of the three internal charges $\Gamma_1$, $\Gamma_2$ and $\Gamma_3$ is an integral charge in $\mathds{Z}^{d(\alpha_d)}\subset {\bf R}_{\alpha_d}$. The $E_d(\mathds{Z})$ invariant\footnote{Here $\mathcal{V}^T = (\mathcal{V}^{-1})^\ddagger$, where $\ddagger$ is the Cartan involution. There always exists a matrix representation of real split groups $E_d(\mathds{R})$ on $R(\Lambda_d)$ such that $\mathcal{V}^T$ is the transpose of the matrix $\mathcal{V}$.}
\be 
g(\Gamma_i,\Gamma_j)=\Gamma_i^T \mathcal{V} \mathcal{V}^T  \Gamma_j  = \frac{1}{2}\bigl( |Z(\Gamma_i+\Gamma_j)|^2 - |Z(\Gamma_i)|^2-|Z(\Gamma_j)|^2\bigr) 
\ee
appearing in the exponential is formed using the coset representative $\mathcal{V}(\Phi)\in E_{d}(\mathds{R})/K(E_d)$, and $|Z(\Gamma)|$ is the mass of a 1/2 BPS state of charge $\Gamma$ satisfying the section constraint $\Gamma\times\Gamma=0$. The integration domain $S^+=(\mathds{R}_+)^6$ denotes the positive Schwinger parameter space. The six Schwinger parameters $L_A$ at three-loop order for the tetrahedral skeleton have been arranged in the symmetric $(3\times 3)$-matrix 
\be 
\Omega =\Omega^{ij} = \left( \begin{array}{ccc} L_1+L_3+L_5   &L_3+L_5 & L_5  \\ L_3+L_5  & L_2+L_3+L_5+L_6  &L_5+L_6 \\ L_5 & L_5+L_6 & L_4+L_5 + L_6 \end{array}\right)  \ . 
\ee
Note that $\Omega$ is a symmetric and positive definite matrix on Schwinger parameter space $S^+$.

There are three internal charges $\Gamma_i$ propagating in the diagram and they all have to mutually satisfy the strong section constraint $\Gamma_i\times \Gamma_j=0$. This is due to the structure of the tetrahedral graph; generally only adjacent charges have to satisfy the section constraint~\cite{Bossard:2015foa}.

Since the $\nabla^6 R^4$ contribution is the lowest contribution from the tetrahedral skeleton there is no dependence on the external momenta and dependence on the Schwinger parameters except for the overall scale and the BPS-mass through the exponential. We have also included an index $\mu$ on the exponential to indicate that the amplitude has to be regulated through the introduction of a mass term $\mu$. We will be more explicit on this regularisation below when we have rewritten the integral in a different form. The integral~\eqref{eq:a1} also exhibits the primitive divergence for $d=5$ (corresponding to six space-time dimensions) that corresponds to the known supergravity 3-loop logarithmic divergence \cite{Bern:2007hh,Bern:2008pv}.

A first step in evaluating~\eqref{eq:a1} consists in showing that the action of $SL(3,\mathds{Z})$ on the symmetric matrices 
\be 
\Omega \rightarrow A \Omega A^T \ , 
\ee 
that preserves Schwinger parameter space $S^+$ generates a finite $S_4$ subgroup of $SL(3,\mathds{Z})$. 
Because the action of $SL(3,\mathds{Z})$ is transitive on $S^+$, any  $SL(3,\mathds{Z})$ transformation that preserves Schwinger parameter space acts by permuting fundamental domains of $SL(3,\mathds{Z})$ in $S^+$. Among all the $S_6$ permutations of the Schwinger parameters, one computes straightforwardly that only a subgroup $S_4\subset S_6$ can be realised in $SL(3,\mathds{Z})$. We have furthermore checked that among hundred thousand $SL(3,\mathds{Z})$ matrices the only ones that preserve $S^+$ all belong to this $S_4$ subgroup, so that there are no other $SL(3,\mathds{Z})$ transformations acting by permutations. One further check consists in computing the integral\footnote{The function $\xi(s)$ appearing here and in many other places in this paper is the completed Riemann zeta function defined by $\xi(s) = \pi^{-s/2} \Gamma(s/2) \zeta(s)$ that satisfies the functional identity $\xi(s)=\xi(1-s)$.} 
\be  
\int_{S^+} \frac{d^6\Omega}{\left(\det \Omega\right)^{2-s}} e^{ -\pi \det \Omega } \approx \frac{24}{\pi^s} \Gamma(s)  \xi(2) \xi(3) =    24 \int_{H^+_{3\times 3}/ SL(3,\mathds{Z})} \frac{d^6\Omega}{\left(\det \Omega\right)^{2-s}} e^{ -\pi \det \Omega } \ ,   
\ee
where we have done the integral on the left numerically on a subset of values for $s\ge 0$. The approximation is such that $24$ is always the closest integer to the resulting value. The integral on the right-hand side contains the space $H^+_{3\times 3}$ of all symmetric positive definite $(3\times 3)$-matrices and the integral is known (see~\eqref{eq:hcos} in the appendix).

One concludes therefore that the subgroup of $SL(3,\mathds{Z})$ that stabilises Schwinger parameter space is $S_4$ of order 24, such that the amplitude~\eqref{eq:a1} reduces to 
\begin{align}
\label{eq:a2}
\mathcal{E}_{\gra{0}{1}}^\ord{\textrm{3-loop}} =20 \sum_{\substack{\Gamma_i\in \mathds{Z}^{3 d(\alpha_d)}\\ \Gamma_i\times \Gamma_j =0}} \int_{H^+_{3\times 3}/SL(3,\mathds{Z})} \frac{d^6\Omega}{\left(\det \Omega\right)^{2-\frac{d-5}{2}}} \exp_\mu\Big( -\pi \Omega^{ij} g(\Gamma_i,\Gamma_j)\Big)\,.
\end{align}

The constrained sum over the three internal charges $\Gamma_i$ can be rewritten by a suitable parabolic decomposition of the $E_{d(d)}$ duality group. This is a generalisation of the discussion in~\cite{Bossard:2015foa} where a similar decomposition was performed at two-loop order. Consider a single charge $\Gamma_1 \in \mathds{Z}^{d(\alpha_d)}$ satisfying $\Gamma_1 \times \Gamma_1=0$, using \cite{Krutelevich}, one has for $d\le 7$ that one can always use an element of the Chevalley subgroup $E_{d}(\mathds{Z})$ to rotate the discrete charge in the highest degree component ${\bf 1}^\ord{10-d}$ in the decomposition 
\begin{align}
\label{1ChargeGrad}  
 \mathfrak{e}_{d(d)} &\cong {\bf R}_{\alpha_{d-1}}^{\ord{d-9}} \oplus \scal{ \mathfrak{gl}_1 \oplus   \mathfrak{e}_{d-1(d-1)}}^\ord{0} \oplus  \overline{\bf R}_{\alpha_{d-1}}^{\ord{9-d}} \ , \CR
{\bf R}_{\alpha_d} &\cong { \updelta}_{d,7}^\ord{-3}  \oplus {\bf R}_{\alpha_1}^\ord{d-8} \oplus {\bf R}_{\alpha_{d-1}}^\ord{1} \oplus {\bf 1}^\ord{10-d} \ ,\CR
{\bf R}_{\alpha_1} &\cong\hspace{5mm}  \dots  \hspace{5mm} \oplus{ \updelta}_{d\ge 6}^\ord{4(d-7)}   \oplus {\bf R}_{\alpha_2}^\ord{d-7} \oplus {\bf R}_{\alpha_1}^\ord{2} \ ,  
\end{align}
Using this one straighforwardly checks that a second charge $\Gamma_2$ such that $\Gamma_1 \times \Gamma_2=0$ must belong to $\overline{\bf R}_{\alpha_{d-1}}^\ord{1} \oplus {\bf 1}^\ord{10-d}$. Using furthermore $\Gamma_2\times \Gamma_2=0$, one can use the same property to conclude that the second charge belongs to the highest level decomposition of $\mathfrak{e}_{d-1(d-1)}$ under $\mathfrak{e}_{d-2(d-2)}$. The two charges can then be chosen to be in the highest degree component ${\bf 2}^\ord{11-d} $ in the decomposition 
\begin{align}
 \label{alphadm1} 
 \mathfrak{e}_{d(d)} &\cong {\bf R}_{\alpha_{1}}^{\ord{2d-18}} \oplus ({\bf 2}\otimes  {\bf R}_{\alpha_{d-2}})^{\ord{d-9}} \oplus \scal{ \mathfrak{gl}_1 \oplus \mathfrak{sl}_2\oplus   \mathfrak{e}_{d-2(d-2)}}^\ord{0} \oplus  ( {\bf 2}\otimes \overline{\bf R}_{\alpha_{d-2}})^{\ord{9-d}} \oplus \overline{\bf R}_{\alpha_{1}}^{\ord{18-2d}}  \ , \CR
{\bf R}_{\alpha_d} &\cong  ( \updelta_{d,7} {\bf 2})^\ord{d-11} \oplus {\bf R}_{\alpha_2}^\ord{2d-16} \oplus ({\bf 2}\otimes  {\bf R}_{\alpha_1})^\ord{d-7} \oplus {\bf R}_{\alpha_{d-2}}^\ord{2} \oplus {\bf 2}^\ord{11-d} \ , \CR
{\bf R}_{\alpha_1} &\cong \hspace{10mm}  \dots  \hspace{10mm}\oplus  ( \updelta_{d,7} ( {\bf 1}\oplus {\bf 3}) \oplus {\bf R}_{\alpha_3})^\ord{2d-14} \oplus ( {\bf 2} \otimes {\bf R}_{\alpha_2})^\ord{d-5} \oplus {\bf R}_{\alpha_1}^\ord{4}  \  . 
\end{align}
Assuming that the two charges $\Gamma_1,\Gamma_2$ are linearly independent, one finds by inspecting the decomposition of the representations above that any third charge $\Gamma_3$ satisfying that $\Gamma_3\times \Gamma_i=0$ must belong to the component  $ \overline{\bf R}_{\alpha_{d-2}}^\ord{2} \oplus {\bf 2}^\ord{11-d} $. Because the component of $\Gamma_3$ in $ \overline{\bf R}_{\alpha_{d-2}}^\ord{2} $ again satisfies the same constraint, one can therefore conclude that the three charges belong to the highest degree component ${\bf 3}^\ord{12-d}$ in the decomposition 
\begin{align}
\mathfrak{e}_{d(d)} &\cong \ldots \oplus \scal{ \mathfrak{gl}_1 \oplus \mathfrak{sl}_3\oplus   \mathfrak{e}_{d-3(d-3)}}^\ord{0} \oplus  ( {\bf 3}\otimes \overline{\bf R}_{\alpha_{d-3}})^{\ord{9-d}} \oplus ( \overline{\bf 3}\otimes \overline{\bf R}_{\alpha_{1}})^{\ord{18-2d}} \oplus \overline{\bf R}_{\alpha_{2}}^\ord{27-3d} \nn\ , \\
{\bf R}_{\alpha_d} &\cong  \ldots\oplus ( {\bf 3} \otimes {\bf R}_{\alpha_2})^\ord{2d-15} \oplus  ( \overline{\bf 3} \otimes {\bf R}_{\alpha_1})^\ord{d-6} \oplus {\bf R}_{\alpha_{d-3}}^\ord{3} \oplus {\bf 3}^\ord{12-d}\ , \nn\\
{\bf R}_{\alpha_1} &\cong \ldots\oplus ( \overline{\bf 3}\otimes {\bf R}_{\alpha_2})^\ord{d-3} \oplus {\bf R}_{\alpha_1}^\ord{6} \ ,
\end{align}
where for $d\leq 6$ the algebra $\mf{e}_{d-3(d-3)}$ has to be interpreted as the correct hidden symmetry obtained from decompactification and the representation ${\bf R}_{\alpha_a}$ has to be interpreted accordingly.\footnote{In particular for $d=4$ one must consider the sum of the two contributions associated to the type IIA and IIB decompactification. For $d=3$ the decompactification is necessarily to eleven-dimensional supergravity for three linearly independent charges, and they cannot be linearly independent for $d<3$.}  The same argument generalises to $\mathfrak{e}_{8(8)}$  using \cite{Bossard:2016hgy}, and one proves in the same way that three charges belonging to the corresponding lattice belong to the highest degree component $ {\bf 3}^\ord{4}$ in
\begin{align}
\mathfrak{e}_{8(8)} &\cong \ldots \oplus \scal{ \mathfrak{gl}_1 \oplus \mathfrak{sl}_3\oplus   \mathfrak{so}(5,5)}^\ord{0} \oplus  ( {\bf 3}\otimes {\bf 16})^{\ord{1}} \oplus ( \overline{\bf 3}\otimes {\bf 10})^{\ord{2}} \oplus \overline{\bf 16}^\ord{3} \oplus {\bf 3}^\ord{4} \nn\ , \\
{\bf 3875} &\cong \ldots\oplus ( \overline{\bf 3}\otimes {\bf 16})^\ord{5} \oplus {\bf 10}^\ord{6} \ . 
\end{align}
The salient point here is that the tensor product of two top components in ${\bf 3}^\ord{12-d}$ always satisfies the strong section constraint (for $d\leq 8$) and is moreover stabilised by the upper parabolic subgroup $P_{d-2}$ with the chosen Levi factor $E_{d-3(d-3)}\times SL(3)\times GL(1)$. We can bring any triplet of charges satisfying the strong section constraint into three copies of the ${\bf 3}^{\ord{12-d}}$, \ie represent them by a $(3\times 3)$-matrix $M$ and conversely any such triplet can be represented as an image of such an $M$ under the action of $E_{d(d)}$ modulo the stabiliser $P_{d-2}$. 

This construction can be understood more generally from the Bruhat decomposition of a Kac--Moody group, and, as a consequence, that any group element defined over $\mathds{Q}$ can be decomposed as the product of an element in the Chevalley group defined over $\mathds{Z}$ and an element in the Borel subgroup over $\mathds{Q}$ , \ie $G(\mathds{Q}) = B(\mathds{Q}) G(\mathds{Z}) $. We discuss this is some details in Appendix \ref{app:E9}, in which we show that the same construction can be generalised to Kac--Moody groups and in particular to $E_{9(9)}$. 

One can therefore rewrite the threshold function~\eqref{eq:a1} as a sum of four terms corresponding to the possible ranks of the matrix $M$
\begin{multline}
\mathcal{E}_{\gra{0}{1}}^\ord{\textrm{3-loop}} = 20\hspace{-5mm}  \int\limits_{H^+_{3\times 3}/SL(3,\mathds{Z})} \frac{d^6\Omega}{\left(\det \Omega\right)^{2-\frac{d-5}{2}}} \left( \sum_{\gamma\in P_{d-2} \backslash E_{d}} \sum_{\substack{M \in \mathds{Z}^{3\times 3}\\ \det M \ne 0}}  \exp\Big( -\pi \Tr (\Omega M \tau^{\scalebox{0.5}{$GL(3)$}}_\gamma M^T)\Big)  \right. \\
+ \sum_{\gamma\in P_{d-1} \backslash E_{d}} \sum_{\substack{M \in \mathds{Z}^{3\times 2}\\ {\rm rk} M =2}}  \exp\Big( -\pi \Tr (\Omega M  \tau^{\scalebox{0.5}{$GL(2)$}}_\gamma M^T)- \pi \mu^2 R_2(\Omega)  \Big)  \\
\left . + \sum_{\gamma\in P_{d} \backslash E_{d}} \sum_{\substack{m \in \mathds{Z}^{3}\\  m \ne 0}}  \exp\Big( -\pi \Tr (\Omega m \tau^{\scalebox{0.5}{$GL(1)$}}_\gamma m^T)- \pi \mu^2 R_1(\Omega) \Big) + \exp\bigl( -\pi \mu^2 R_0(\Omega)\bigr)  \right) 
\end{multline}
Here, $\tau^{\scalebox{0.5}{$GL(n)$}}_\gamma$ is the $GL(n)$ symmetric matrix representating the $E_{d(d)}$ representative in the top component ${\bf n}^\ord{9-d+n}$ in the decomposition of ${\bf R}_{\alpha_d}$, after the action of the discrete $\gamma\in P_{d+1-n} \backslash E_{d}(\mathds{Z})$ coset representative $\tau^{\scalebox{0.5}{$GL(n)$}}_\gamma(\mathcal{V}) = \tau^{\scalebox{0.5}{$GL(n)$}}(\gamma \mathcal{V})$. Note indeed that  the representatives of the charges only contract into this part of the full $E_{d}(\mathds{R})$ coset element, such that with the corresponding embedding of $M \in \mathds{Z}^{ n\times d(\alpha_d)}$, $M\tau^{\scalebox{0.5}{$GL(n)$}}(\mathcal{V}) M^T = M \mathcal{V} \mathcal{V}^T M^T$. $R_n(\Omega)$ are some functions of $\Omega$ that we shall specify below, that regulate the infrared divergence with the infrared regulating mass $\mu$. The specific contributions in $\log\mu$ that will be relevant in the following should not depend on the specific choice of function $R_n(\Omega)$, so we shall choose them such as to make the computation as simple as possible. 

We will now unfold the integral over $H^+_{3\times 3}/SL(3,\mathds{Z})$ for each orbit in the above equation. For non-degenerate three by three matrices the stabiliser of $SL(3,\mathds{Z})$ is trivial so that we can unfold the integral to $H^+_{3\times 3}$. For simplicity we shall consider twice the sum over non-degenerate matrices in 
$\mathds{Z}^{3\times 3}/GL(3,\mathds{Z})$ (rather than once the matrices in $\mathds{Z}^{3\times 3}/SL(3,\mathds{Z})$).

For rank two three by two matrices, the stabiliser is $\mathds{Z}^2 \subset SL(3,\mathds{Z})$, and $SL(3,\mathds{Z})$ allows to rotate $M$ to two by two representatives of non-vanishing determinant in $\mathds{Z}^{2\times 2} /  GL(2,\mathds{Z})$. This choice distinguishes the decomposition of  the symmetric matrix of Schwinger parameters in block form as
\be
\Omega = \begin{pmatrix}\Omega_{2\times 2} & \Omega_{2\times 2} u\\ 
u^T  \Omega_{2\times 2} &u^T \Omega_{2\times 2}u +t \end{pmatrix}
\,\Rightarrow\,\det\Omega = t\, \det\Omega_{2\times 2} \ , 
\ee
and
\be 
d^6\Omega = \left(\det \Omega_{2\times 2} \right) d^3\Omega_{2\times 2}\, d^2u\, dt \,, 
\ee
such that $\Tr (\Omega_{3\times 3} M_{3\times2}  \tau^{\scalebox{0.5}{$GL(2)$}}_\gamma M^T_{3\times 2}) = \Tr (\Omega_{2\times 2} M_{2\times 2}  \tau^{\scalebox{0.5}{$GL(2)$}}_\gamma M^T_{2\times 2})$, and the $\mathds{Z}^2$ stabiliser acts as a shift of $u$. Choosing for convenience $R_{2}(\Omega) = t$ to regularise the integral, the integral of $u$ simply gives a unit volume contribution. 

For non-zero vectors $m\in \mathds{Z}^3$ the stabiliser is $SL(2,\mathds{Z})\ltimes \mathds{Z}^2 \subset SL(3,\mathds{Z})$, and $SL(3,\mathds{Z})$ permits to rotate $m$ to a positive integer in the first component. This choice distinguishes the decomposition of  the symmetric matrix of Schwinger parameters in block form as
\be
\Omega = \begin{pmatrix}\Omega_{1\times 1} & \Omega_{1\times 1} u^T\\ 
u \Omega_{1\times 1} &\, t_{2\times 2}+u \Omega_{1\times 1} u^T  \end{pmatrix}
\,\Rightarrow\,\det\Omega = \det t_{2\times 2} \, \Omega_{1\times 1} \ , 
\ee
and
\be d^6\Omega = \left(\det t_{2\times 2} \right) d\Omega_{1\times 1}\, d^2u\, d^3t_{2\times 2} \,, \ee
such that $\Tr (\Omega_{3\times 3} m  \tau^{\scalebox{0.5}{$GL(1)$}}_\gamma m^T) = \Omega_{1\times 1}  \tau^{\scalebox{0.5}{$GL(1)$}}_\gamma m^2$, and the $SL(2,\mathds{Z}) \ltimes \mathds{Z}^2$ stabiliser acts as a shift of $u$, and linearly on $u$ and $t$. Choosing for convenience $R_{1}(\Omega) = \det t_{2\times 2}$ to regularise the integral, the integral of $u$ simply gives a $\frac{1}{2}$ volume contribution (because of the $-\mathds{1} \in SL(2,\mathds{Z})$ that does not act on $t$). The remaining integral over the matrix $t_{2\times 2}$ is over ${H^+_{2\times 2}/PSL(2,\mathds{Z})}$, which is twice the integral over ${H^+_{2\times 2}/PGL(2,\mathds{Z})}$, so we reabsorb the factor of $\frac{1}{2}$ of the integral over $u$ by this halving of the integration domain of $t_{2\times 2}$.

For the trivial orbit the stabiliser is of course $SL(3,\mathds{Z}) = PGL(3,\mathds{Z})$. For convenience we write the variable as $t_{3\times 3}$, since the integral contribution is defined by its infrared divergence, and we use  $R_{0}(t_{3\times 3}) = \det t_{3\times 3}$ for simplicity. 

So to conclude, the orbit method permits to reduce the threshold function to
\begin{align}
\label{eq:a3}
\mathcal{E}_{\gra{0}{1}}^\ord{\textrm{3-loop}} &=40 \sum_{\gamma\in P_{d-2} \backslash E_{d}} \sum_{\substack{M \in \mathds{Z}^{3\times 3}/GL(3,\mathds{Z})\\ \det(M)\neq 0} } \,\,\int\limits_{H^+_{3\times 3}} \frac{d^6\Omega}{\left(\det \Omega\right)^{2-\frac{d-5}{2}}} \exp\Big( -\pi \Tr (\Omega M \tau^{\scalebox{0.5}{$GL(3)$}}_\gamma M^T)\Big)\nn\\
&\quad +20  \sum_{\gamma\in P_{d-1} \backslash E_{d}} \sum_{\substack{M \in \mathds{Z}^{2\times 2}/GL(2,\mathds{Z})\\ \det(M)\neq 0} } \,\,\int\limits_{H^+_{2\times 2}} \frac{d^3\Omega}{\left(\det \Omega\right)^{\frac32-\frac{d-4}{2}}}\int\limits_0^\infty \frac{dt}{t^{1-\frac{d-7}{2}}} \nn\\
&\hspace{30mm} \times  \exp\Big( -\pi \Tr (\Omega M \tau^{\scalebox{0.5}{$GL(2)$}}_\gamma  M^T)-\pi  \mu^2 t \Big)\nn\\
&\quad +20  \sum_{\gamma\in P_{d} \backslash E_{d}}  \sum_{m>0} \,\,\int\limits_{0}^\infty \frac{d\Omega}{\Omega^{1-\frac{d-3}{2}}}\int\limits_{H^+_{2\times 2}/PGL(2,\mathds{Z})} \frac{d^3t}{\left(\det t\right)^{\frac32-\frac{d-6}{2}}} \nn\\
&\hspace{30mm} \times  \exp\Big( -\pi  \Omega   \tau^{\scalebox{0.5}{$GL(1)$}}_\gamma  m^2 -\pi  \mu^2 \det t \Big)\nn\\
&\quad + 20 \int\limits_{H^+_{3\times 3}/PGL(3,\mathds{Z})} \frac{d^6t}{\left(\det t\right)^{2-\frac{d-5}{2}}}   \exp\Big(  -\pi \mu^2  \det t  \Big) \,.
\end{align}
In this formula we have suppressed the subscripts on the sub-blocks in $\Omega$,  $M$ and $t$ in order to ease the notation. Their size is evident from the summation and integration ranges.

In the next step, we carry out the integrals over $\Omega$ and $t$ using the formul\ae{} of Appendix~\ref{app:matrixint}. These reduce the expressions for each rank into a power of $\det\tau_\gamma$ multiplied by a power of the regulator $\mu$ and $d$-dependent numerical factors involving $\Gamma$ and $\xi$ factors. The result is
\begin{align}
\mathcal{E}_{\gra{0}{1}}^\ord{\textrm{3-loop}} &=40 \xi(d-5)\xi(d-6)\xi(d-7)\sum_{\gamma\in P_{d-2} \backslash E_{d}} (\det \tau^{\scalebox{0.5}{$GL(3)$}}_\gamma)^{-\frac{d-5}2}\nn\\
&\quad+ 20 \Gamma\left(\frac{d-7}2\right) (\pi\mu^2)^{-\frac{d-7}2} \xi(d-4)\xi(d-5) \sum_{\gamma\in P_{d-1} \backslash E_{d}} (\det\tau^{\scalebox{0.5}{$GL(2)$}}_\gamma)^{-\frac{d-4}2}\nn\\
&\quad+ 20 \xi(2) \Gamma\left(\frac{d-6}2\right) (\pi\mu^2)^{-\frac{d-6}2} \xi(d-3) \sum_{\gamma\in P_{d} \backslash E_{d}}(\tau^{\scalebox{0.5}{$GL(1)$}}_\gamma)^{-\frac{d-3}2}\nn\\
&\quad +20 \xi(2)\xi(3) \Gamma\left(\frac{d-5}2\right) (\pi\mu^2)^{-\frac{d-5}2}\,.
\end{align}
The $\gamma$-sums over the duality group can be carried out to yield the final result
\begin{align}
\label{eq:a4}
\mathcal{E}_{\gra{0}{1}}^\ord{\textrm{3-loop}} &=40 \xi(d-5)\xi(d-6)\xi(d-7) E_{\alpha_{d-2},\frac{d-5}2}\nn\\
&\quad  + 20 \Gamma\left(\frac{d-7}2\right) (\pi\mu^2)^{-\frac{d-7}2} \xi(d-4)\xi(d-5) E_{\alpha_{d-1},\frac{d-4}2}\nn\\
&\quad+ 20 \xi(2) \Gamma\left(\frac{d-6}2\right) (\pi\mu^2)^{-\frac{d-6}2} \xi(d-3) E_{\alpha_{d},\frac{d-3}2}\nn\\
&\quad+20 \xi(2)\xi(3) \Gamma\left(\frac{d-5}2\right) (\pi\mu^2)^{-\frac{d-5}2}\,,
\end{align}
where the Langlands Eisenstein series coming from the maximal parabolic cosets sums have been labelled by the node associated with the maximal parabolic subgroup of $E_{d(d)}$ together with the parameter of the inducing determinant.

Before explaining the derivation~\eqref{eq:a4} in more detail, we make a small parenthesis on our different conventions for denoting Eisenstein series. More precisely, we have used for $1\le n\le 3$
\begin{align}
E_{\alpha_{d+1-n},s} = \sum_{\gamma\in P_{d+1-n} \backslash E_{d}} \left( \det \tau_\gamma^{GL(n)} \right)^{-s}\,, 
\end{align}
such that the identity coset term has numerical coefficient equal to one.  We shall also encounter Eisenstein series associated with non-maximal parabolic subgroups and in this case it is convenient to either label the series by putting the corresponding weight $\{ s_i\}_{i=1}^d$ on the Dynkin diagram (for a fixed symmetry group $E_{d(d)}$) or by writing the weight $\sum_{i=1}^d s_i \Lambda_i $ in the basis of the fundamental weights $\{ \Lambda_i \}_{i=1}^d$. Note that the Eisenstein series are instead commonly labeled by the weight $\lambda = 2 \sum_{i=1}^d s_i \Lambda_i - \rho =\sum_{i=1}^d (2s_i-1) \Lambda_i  $ defining the infinitesimal character on which the Weyl group acts in functional relations. For ease of notation it will be nonetheless useful to label them by $ \sum_{i=1}^d s_i \Lambda_i $ for short, since most of the $s_i$ vanish in practice. Concretely, we write for maximal parabolic Eisenstein series  
\begin{align}
\label{eq:ESconv}
E_{\alpha_{i},s} = E_{s \Lambda_{i}} 
\end{align}
indicating that the fundamental weight $\Lambda_{i}$ occurs with coefficient $s$ in the weight. Moreover, we use the $E_d$ labelling for the fundamental weights and the standard labelling for the Dynkin diagrams, such that we would write for example for $E_{5(5)}= SO(5,5)$ of type $D_5$
\begin{align}
E_{\sum_{i=1}^d s_i \Lambda_i } = E_{\mbox{\DSOX{\mathnormal{s_1}}{\mathnormal{s_3}}{\mathnormal{s_2}}{\mathnormal{s_4}}{\mathnormal{s_5}}}}\,.
\end{align}
In deriving the final expression~\eqref{eq:a4}, we have used the identities for the matrix integrals and Eisenstein series naively and without paying attention to their convergence. In fact one can check that the integral over $t$ only converges absolutely for the rank $n$ orbit if $d>5+n$ and the integral over $\Omega$  for the rank $n$ orbit if $d>1+2n$. To take care of the convergence of the Langlands Eisenstein series we consider the analytic continuation of the parameter by replacing $d$ by $d+2\epsilon$, which corresponds formally to dimensional regularisation. In this case one checks that the Eisenstein series $E_{\scalebox{0.6}{$\alpha_{d+1-n},\frac{d-2-n}{2}+\epsilon$}}$ converges absolutely for $\mbox{Re}(\epsilon) >  \frac{n^2-(d-2)n+7d-3}{2(9+n-d)}$, using the convergence criterion that $E_{s \Lambda_i}$ is absolutely convergent if and only if $\langle \Lambda_i , s \Lambda_i - \rho\rangle >0$ for a maximal parabolic Eisenstein series.\footnote{The case $d=5$ and $n=3$ has to be treated separately since it is not a maximal parabolic series and we will give its convergence condition below after~\eqref{eq:D5}.} These expression are therefore generally divergent at $\epsilon=0$, but are absolutely convergent for $\mbox{Re}(\epsilon)$ satisfying the above inequality. Using Langlands' construction these functions can then be analytically extended to meromorphic functions in $\epsilon$ to the whole complex plane, where one also continues the numerical prefactors appropriately. We shall use this analytic continuation as a dimensional regularisation as in~\cite{Bossard:2015foa}. Each individual expression is then finite for a dense set of $\epsilon\in \mathds{C}$, but the expressions involve individually poles at $\epsilon=0$. These divergences are also to be expected on physical grounds and signal the appearance of ultra-violet divergences of amplitudes or form factors in supergravity, or equivalently, ambiguities in the decomposition of the non-perturbative string amplitude into analytic and non-analytic components due to the logarithmic behaviour of the latter in the Mandelstam variables.  We will now discuss the treatment of these divergences in the different dimensions, and show that all poles cancel in the complete expression for the amplitude for all $d$.

\subsection{$D=6$}

For six space-time dimensions ($d=5$) one has to interpret the Eisenstein series from the various orbits as follows where we also introduce the dimensional regularisation $d=5+2\epsilon$:
\begin{align}
\label{eq:D5}
E_{\alpha_{d-2},\epsilon} = E_{\mbox{\DSOX{0}{\epsilon}{\epsilon}{0}{0}}}\,,\quad
E_{\alpha_{d-1},\scalebox{0.6}{$\frac12$}+\epsilon} = E_{\mbox{\DSOX{0}{0}{0}{\mathnormal{\Tfrac12\mbox{+}\epsilon}}{0}}}\,,\quad
E_{\alpha_{d},1+\epsilon} = E_{\mbox{\DSOX{0}{0}{0}{0}{\mathnormal{1\mbox{+}\epsilon}}}}\,.
\end{align}
In particular, the series $E_{\alpha_{d-2},\epsilon}$ is not a maximal parabolic Eisenstein series and converges absolutely for $\mbox{Re}(\epsilon)>\tfrac52$.  Considering the factors multiplying the various terms we see that the contributions from the rank-one orbit and rank-two orbit give only finite contributions for $\epsilon\to 0$ that vanish when sending the IR regulator $\mu$ to zero. The rank-three and rank-zero orbits on the other hand give divergent contributions that we now analyse.\footnote{For the Eisenstein series themselves one has $E_{\alpha_d,1+\epsilon}^{SO(5,5)}=O(\epsilon^0)$, $E_{\alpha_{d-1},\Tfrac12+\epsilon}^{SO(5,5)}=O(\epsilon^2)$ and $E_{\alpha_{d-2},\epsilon}^{SO(5,5)}=O(\epsilon^0)$.}

Let us first analyse the series coming from the non-degenerate orbit. This term is divergent due to the $\xi(d-5)\rightarrow \xi(2\epsilon)$ prefactor. To analyse it, we note the following functional relations of Eisenstein series 
\begin{align}
E_{\mbox{\DSOX0{\epsilon}000}} &= \frac{\xi(5-2\epsilon)\xi(6-2\epsilon)\xi(7-2\epsilon)\xi(8-4\epsilon)}{\xi(1-2\epsilon)\xi(2-2\epsilon)\xi(3-2\epsilon)\xi(7-4\epsilon)} E_{\mbox{\DSOX0{\mathnormal{\Tfrac72\mbox{-}\epsilon}}000}}\,,\nn\\
E_{\mbox{\DSOX00{\epsilon}00}} &= \frac{\xi(6-2\epsilon)\xi(8-2\epsilon)}{\xi(1-2\epsilon)\xi(3-2\epsilon)} E_{\mbox{\DSOX0000{\mathnormal{4\mbox{-}\epsilon}}}} \,.
\end{align}
Using the property that 
\be E_{\mbox{\DSOX{0}{\epsilon}{\epsilon}{0}{0}}}   = E_{\mbox{\DSOX{0}{\epsilon}{0}{0}{0}}} +E_{\mbox{\DSOX{0}{0}{\epsilon}{0}{0}}} -E_{\mbox{\DSOX{0}{0}{0}{0}{0}}} + \mathcal{O}(\epsilon^2) \ , \ee
one deduces therefore that the rank-three contribution becomes
\begin{align}
\label{eq:ND5}
\xi(2)\xi(3)\xi(2\epsilon) E_{\mbox{\DSOX{0}{\epsilon}{\epsilon}{0}{0}}} =  \xi(2)\xi(3)\xi(2\epsilon) +\xi(5)\xi(6)\xi(8) \hat{E}_{\mbox{\DSOX0{\mathnormal{\Tfrac72}}000}} + \xi(2)\xi(6)\xi(8)  \hat{E}_{\mbox{\DSOX0000{\mathnormal{4}}}} +O(\epsilon)\,,
\end{align}
where the hat on the Eisenstein series indicates that the pole in $\frac{1}{\epsilon}$ has been removed before taking the limit $\epsilon\to 0$, \eg
\begin{align}
  \hat{E}_{\mbox{\DSOX0000{\mathnormal{4}}}}  &= \lim_{\epsilon\rightarrow 0} \biggl( E_{\mbox{\DSOX0000{\mathnormal{4\! +\! \epsilon}}}}  - \frac{\xi(1+2\epsilon)\xi(3+2\epsilon)}{\xi(6+2\epsilon)\xi(8+2\epsilon)} \biggr) \,, \nn\\
 \hat{E}_{\mbox{\DSOX0{\mathnormal{\Tfrac72}}000}}  &= \lim_{\epsilon\rightarrow 0} \biggl(  {E}_{\mbox{\DSOX0{\mathnormal{\Tfrac72\! +\! \epsilon}}000}}  - \frac{\xi(1+2\epsilon)\xi(2+2\epsilon)\xi(3+2\epsilon)\xi(7+4\epsilon)}{\xi(5+2\epsilon)\xi(6+\epsilon)\xi(7+2\epsilon)\xi(8+4\epsilon)}   \biggr) \,.
  \end{align}
The remaining explicit divergence in~\eqref{eq:ND5} associated with $\xi(2\epsilon)\sim -\frac{1}{2\epsilon}$, cancels precisely the leading part in the IR divergence in the last term of~\eqref{eq:a4} coming from the rank-zero orbit. 

Putting everything together we therefore obtain (for some irrelevant constant $c_1$)
\begin{align}
\label{eq:6DE3}
\mathcal{E}_{\gra{0}{1}}^\ord{\textrm{3-loop}} = 40 \xi(5)\xi(6)\xi(8) \hat{E}_{\mbox{\DSOX0{\Tfrac72}000}} + 40 \xi(2)\xi(6)\xi(8)  \hat{E}_{\mbox{\DSOX0000{4}}}  -20 \xi(2)\xi(3) (\log(\pi \mu^2)+c_1)  \,.
\end{align}
The two series are the two homogeneous solutions to the differential equation for the $\nabla^6R^4$ term and the combination is the one displayed in~\cite{Bossard:2015oxa}. The constant logarithmic term in $\xi(2)\xi(3) \log(\pi \mu^2)$ exhibits the need of introducing a renormalisation scale in the non-analytic component of the amplitude, which is a consequence of the logarithmic divergence in the supergravity four-graviton scattering amplitude at 3-loop \cite{Bern:2008pv}. 

\subsection{$D=5$}

For five space-time dimensions ($d=6+2\epsilon$) one has to perform a similar analysis to above. The only interesting terms are the rank-three and the rank-one orbit in this case
\begin{align}
\label{eq:a54D}
\mathcal{E}_{\gra{0}{1}}^\ord{\textrm{3-loop}} &=40 \xi(2-2\epsilon)\xi(1+2\epsilon)\xi(2\epsilon)E_{\mbox{{{\tiny $ { \left[ \begin{array}{cccccc}  & & \mathfrak{0} \hspace{-0.7mm}&&& \vspace{ -1mm} \\ \mathfrak{0}\hspace{-0.7mm} &  \mathfrak{0} \hspace{-0.7mm}&\Tfrac12\mbox{+}\epsilon \hspace{-0.7mm} & \mathfrak{0}\hspace{-0.7mm}& \mathfrak{0} \end{array}\right] }$}}}} + 20 \xi(2) \Gamma(\epsilon) ( \pi \mu^2)^{-\epsilon} \xi(3+2\epsilon)E_{\mbox{\DEVI00000{\Tfrac32\mbox{+}\epsilon}}}  \,.
\end{align}
Using the functional relations 
\begin{align}
\xi(1+2\epsilon) E_{\mbox{{{\tiny $ { \left[ \begin{array}{cccccc}  & & \mathfrak{0} \hspace{-0.7mm}&&& \vspace{ -1mm} \\ \mathfrak{0}\hspace{-0.7mm} &  \mathfrak{0} \hspace{-0.7mm}&\Tfrac12\mbox{+}\epsilon \hspace{-0.7mm} & \mathfrak{0}\hspace{-0.7mm}& \mathfrak{0} \end{array}\right] }$}}}} 
&= \xi(3-2\epsilon) E_{\mbox{\DEVI0{\mathnormal{\epsilon}}{\mathnormal{\epsilon}}00{\mathnormal{\Tfrac32\mbox{-}\epsilon}}}} \,, \\
E_{\mbox{{{\tiny $ { \left[ \begin{array}{cccccc}  & &{\mathnormal{\mbox{-}\epsilon}}  \hspace{-0.7mm}&&& \vspace{ -1mm} \\\mathfrak{0} \hspace{-0.7mm} &  \mathfrak{0} \hspace{-0.7mm}&\mathfrak{0} \hspace{-0.7mm} & \mathfrak{0}\hspace{-0.7mm}& {\mathnormal{\Tfrac32\mbox{+}\epsilon}}\end{array}\right] }$}}}} 
 &= \frac{\xi(6+2\epsilon)\xi(9+2\epsilon)}{\xi(1+2\epsilon)\xi(3+2\epsilon)} E_{\mbox{\DEVI00000{\mathnormal{\Tfrac92\mbox{+}\epsilon}}}} \, , \nn\\
E_{\mbox{\DEVI00{\epsilon}00{\mathnormal{\Tfrac32\mbox{-}\epsilon}}}} &= \frac{\xi(2+2\epsilon)}{\xi(3-2\epsilon)} E_{\mbox{{{\tiny $ { \left[ \begin{array}{cccccc}  & &\hspace{-1mm} 1\mbox{+} \epsilon\hspace{-1mm} \hspace{-0.7mm}&&& \vspace{ -1mm} \\\mathfrak{0} \hspace{-0.7mm} &  \mathfrak{0} \hspace{-0.7mm}&\mathfrak{0} \hspace{-0.7mm} & \mathfrak{0}\hspace{-0.7mm}& \mathfrak{0} \end{array}\right] }$}}}}=  \frac{\xi(6-2\epsilon)\xi(7-2\epsilon)\xi(8-4\epsilon)\xi(9-2\epsilon)}{\xi(2\epsilon)\xi(2-2\epsilon)\xi(3-2\epsilon)\xi(7-4\epsilon)} E_{\mbox{{{\tiny $ { \left[ \begin{array}{cccccc}  & &\hspace{-1mm} \frac92\mbox{-} \epsilon\hspace{-1mm} \hspace{-0.7mm}&&& \vspace{ -0.5mm} \\\mathfrak{0} \hspace{-0.7mm} &  \mathfrak{0} \hspace{-0.7mm}&\mathfrak{0} \hspace{-0.7mm} & \mathfrak{0}\hspace{-0.7mm}& \mathfrak{0} \end{array}\right] }$}}}} \nn
 \, , 
\end{align}
one can simplify the rank-three orbit contribution similarly as for~\eqref{eq:ND5}
\begin{multline}
\label{eq:ND6}
\xi(2\epsilon)\xi(1+2\epsilon) E_{\mbox{{{\tiny $ { \left[ \begin{array}{cccccc}  & & \mathfrak{0} \hspace{-0.7mm}&&& \vspace{ -1mm} \\ \mathfrak{0}\hspace{-0.7mm} &  \mathfrak{0} \hspace{-0.7mm}&\Tfrac12\mbox{+}\epsilon \hspace{-0.7mm} & \mathfrak{0}\hspace{-0.7mm}& \mathfrak{0} \end{array}\right] }$}}}}  
= \xi(2+2\epsilon) \xi(2\epsilon) 
E_{\mbox{{{\tiny $ { \left[ \begin{array}{cccccc}  & &\hspace{-1mm} 1\mbox{+} \epsilon\hspace{-1mm} \hspace{-0.7mm}&&& \vspace{ -1mm} \\\mathfrak{0} \hspace{-0.7mm} &  \mathfrak{0} \hspace{-0.7mm}&\mathfrak{0} \hspace{-0.7mm} & \mathfrak{0}\hspace{-0.7mm}& \mathfrak{0} \end{array}\right] }$}}}} 
+\xi(3-2\epsilon) \xi(2\epsilon) E_{\mbox{\DEVI00000{\mathnormal{\Tfrac32\mbox{+}\epsilon}}}}\\-\frac{\xi(2\epsilon) \xi(3-2\epsilon)}{ \xi(-2\epsilon) \xi(3+2\epsilon)} \xi(6+2\epsilon)\xi(9+2\epsilon) E_{\mbox{\DEVI00000{\mathnormal{\Tfrac92\mbox{+}\epsilon}}}}
+O(\epsilon)\,.
\end{multline}
The divergent second term in~\eqref{eq:ND6} cancels against the rank-one contribution, leaving only finite pieces and logarithms of the IR regulator $\mu$. With a similar appropriate definition of the regularised Eisenstein series one obtains 
\be 
\label{eq:5DE3}
\mathcal{E}_{\gra{0}{1}}^\ord{\textrm{3-loop}}  = 40 \xi(6) \xi(8)\xi(9) \hat{E}_{\mbox{{{\tiny $ { \left[ \begin{array}{cccccc}  & &\hspace{-1mm} \Tfrac92 \hspace{-1mm} \hspace{-0.7mm}&&& \vspace{-0.5mm} \\\mathfrak{0} \hspace{-0.7mm} &  \mathfrak{0} \hspace{-0.7mm}&\mathfrak{0} \hspace{-0.7mm} & \mathfrak{0}\hspace{-0.7mm}& \mathfrak{0} \end{array}\right] }$}}}} +  40 \xi(2) \xi(6) \xi(9) \hat{E}_{\mbox{\DEVI00000{\mathnormal{\Tfrac92}}}} - 20 \xi(2) \xi(3) \log( \pi \mu^2) E_{\mbox{\DEVI00000{\Tfrac32}}} \ . 
\ee
Once again, the two  series  defining the $\mu$ independent contribution are the two homogeneous solutions to the differential equation for the $\nabla^6R^4$ term and the combination is the one displayed in~\cite{Bossard:2015oxa}. The constant logarithmic term in $\xi(2) \xi(3) \log( \pi \mu^2)$ exhibits the need of introducing a renormalisation scale in the non-analytic component of the amplitude, which is a consequence of the logarithmic divergence in the supergravity form factor of the $\mathcal{E}_{\gra{0}{0}} R^4$ type invariant with four external gravitons at 2-loop.

\subsection{$D=4$}
For $d=7+ 2 \epsilon$, it is the rank 2 orbit that is divergent in the limit $\mu\rightarrow 0$, and one gets together with the rank 3 orbit
\begin{align}
\label{eq:a44D}
\mathcal{E}_{\gra{0}{1}}^\ord{\textrm{3-loop}} &=40 \xi(2+2\epsilon)\xi(1+2\epsilon)\xi(2\epsilon)E_{\mbox{\DEVII0000{\mathnormal{1\mbox{+}\epsilon\, }}0{\mathfrak{0}}}}+ 20 \Gamma(\epsilon) ( \pi \mu^2)^{-\epsilon} \xi(2+2\epsilon) \xi(3+ 2 \epsilon) E_{\mbox{\DEVII00000{\mathnormal{\Tfrac{3}{2}\mbox{+}\epsilon\, }}{\mathfrak{0}}}}  \,,
\end{align}
Using successive Weyl group transformations one proves the identities 
\begin{align}
E_{\mbox{\DEVII0000{\mathnormal{1\mbox{+}\epsilon\, }}0{\mathfrak{0}}}}  = \frac{\xi(3-2\epsilon)\xi(4-2\epsilon)\xi(7-4\epsilon)\xi(8-6\epsilon)}{\xi(1+2\epsilon)\xi(2+2\epsilon)\xi(3-4\epsilon)\xi(7-6\epsilon)} E_{\mbox{\DEVII00{\mathnormal{\epsilon}}00{\mathnormal{\epsilon\, }}{{4\mbox{-}3\epsilon}}}}
\end{align}
and 
\begin{align}
 E_{\mbox{\DEVII00000{\mathnormal{\epsilon}}{{4\mbox{-}2\epsilon}}}}  &=  \frac{\xi(9-2\epsilon)\xi(12-2\epsilon)}{\xi(1-2\epsilon)\xi(4-2\epsilon)} E_{\mbox{\DEVII{\mathnormal{6\mbox{-}\epsilon}}00000{\mathfrak{0}}}} \ , \CR
E_{\mbox{\DEVII00{\mathnormal{\epsilon}}000{{4\mbox{-}2\epsilon}}}}  &= \frac{\xi(2+2\epsilon)\xi(3+2\epsilon)\xi(3-4\epsilon)}{\xi(4-2\epsilon)\xi(3-2\epsilon)\xi(8-4\epsilon)} E_{\mbox{\DEVII00000{\mathnormal{\Tfrac{3}{2}\mbox{+}\epsilon\, }}{\mathfrak{0}}}} \ , \label{FunctionalE7various4}
\end{align} 
which imply that 
\begin{multline}
E_{\mbox{\DEVII00{\mathnormal{\epsilon}}00{\mathnormal{\epsilon\, }}{{4\mbox{-}3\epsilon}}}} = \frac{\xi(9-2\epsilon)\xi(12-2\epsilon)}{\xi(1-2\epsilon)\xi(4-2\epsilon)} E_{\mbox{\DEVII{\mathnormal{6\mbox{-}\epsilon}}00000{\mathfrak{0}}}} 
+ \frac{\xi(2+2\epsilon)\xi(3+2\epsilon)\xi(3-4\epsilon)}{\xi(4-2\epsilon)\xi(3-2\epsilon)\xi(8-4\epsilon)} E_{\mbox{\DEVII00000{\mathnormal{\Tfrac{3}{2}\mbox{+}\epsilon\, }}{\mathfrak{0}}}} \\
- E_{\mbox{\DEVII000000{\, 4\mbox{-}\epsilon}}} +\mathcal{O}(\epsilon^2)\,.
\end{multline}
Substituting this last expression in \eqref{eq:a44D}, one gets that the poles in $E_{\mbox{\DEVII00000{\mathnormal{\Tfrac{3}{2}}}{\mathfrak{0}}}}$ cancel between the rank three and the rank two orbits, and using \eqref{FunctionalE7various4} that the poles in $E_{\mbox{\DEVII000000{4}}}$ cancel as well, such that the final expression is finite. Using moreover the identities 
\be \xi(8-2\epsilon) \xi(4-2\epsilon)\xi(-2\epsilon) E_{\mbox{\DEVII000000{\, 4\mbox{-}\epsilon}}} = \xi(2+2\epsilon) \xi(6+2\epsilon) \xi(10+2\epsilon) E_{\mbox{\DEVII000000{\, 5\mbox{+}\epsilon}}} \ , \ee
and 
\be  \xi(3) E_{\mbox{\DEVII00000{\mathnormal{\Tfrac{3}{2}}}{\mathfrak{0}}}} =  \xi(5) E_{\mbox{\DEVII{\mathnormal{\Tfrac52}}00000{\mathfrak{0}}}} \;, \ee
one obtains finally 
\be
\label{4DE3}
\mathcal{E}_{\gra{0}{1}}^\ord{\textrm{3-loop}} =40 \xi(8) \xi(9) \xi(12) \hat{E}_{\mbox{\DEVII{\mathnormal{6}}00000{\mathfrak{0}}}} + 40 \xi(2) \xi(6) \xi(10) \hat{E}_{\mbox{\DEVII000000{5}}} - 20 \xi(2) \xi(5) \log( \pi \mu^2 ) E_{\mbox{\DEVII{\mathnormal{\Tfrac52}}00000{\mathfrak{0}}}} \ , \ee
where the hatted functions are defined to be the finite part of the corresponding divergent Eisenstein series, for which the pole $ ( \tfrac{1}{\epsilon} + c_E) E{\mbox{\DEVII{\mathnormal{\Tfrac52}}00000{\mathfrak{0}}}}$ has been removed for a given choice of constant $c_E$ . Here we do not define a precise subtraction scheme (defining $c_E$), since this would only become meaningful if we were  also considering the appropriately regularised non-analytic part of the amplitude such that the complete amplitude would match correctly the perturbative string theory three-loop amplitude. The full 3-loop non-analytic amplitude is not known. The two  series  defining the $\mu$ independent contribution are the two homogeneous solutions to the differential equation for the $\nabla^6R^4$ term and the combination is the one displayed in~\cite{Bossard:2015oxa}. The constant logarithmic term in $\xi(2) \xi(5) \log( \pi \mu^2)$ exhibits the need of introducing a renormalisation scale in the non-analytic component of the amplitude, which is a consequence of the logarithmic divergence in the supergravity form factor of the $\mathcal{E}_{\gra{1}{0}} \nabla^4 R^4$ type invariant with four extrenal gravitons at 1-loop in four dimensions.

\subsection{$D=3$}

For $D=3$ ($d=8$) there is not much to do. The non-degenerate orbit satisfies the functional relation
\begin{align}
E_{\mbox{\DEVIII00000{\Tfrac32\!+\!\epsilon}00}} = \frac{\xi(2-2\epsilon)\xi(3-2\epsilon)\xi(4-2\epsilon)\xi(6-4\epsilon)\xi(7-4\epsilon)\xi(11-6\epsilon)}{\xi(1+2\epsilon)\xi(2+2\epsilon)\xi(3+2\epsilon)\xi(3-4\epsilon)\xi(4-4\epsilon)\xi(7-6\epsilon)} E_{\mbox{\DEVIII000\epsilon000{\mathnormal{\Tfrac{1\! 1}2\mbox{-} 3\epsilon}}}}\ , 
\end{align}
and so directly relates in a regular way to the adjoint function at $s_{8} = \frac{11}2$ that solves the homogeneous supesrymmetry differential equations and that comes with a finite overall coefficient. In dimensions $D>3$ we always had combinations of adjoint and fundamental Eisenstein series. For $E_8$, these two notions coincide and that is why the presence of only one function here agrees with the expectations. The two classes of supersymmetry invariants with couplings satisfying two distinct sets of differential equations also coincide in $D=3$ and there is a unique class of supersymmetry invariant \cite{Bossard:2015uga}.

The final answer obtained from our 3-loop calculation is then
\be
\label{3DE3}
\mathcal{E}_{\gra{0}{1}}^\ord{\textrm{3-loop}} =40 \xi(8) \xi(9) \xi(11) E_{\mbox{\DEVIII0000000{\mathnormal{\Tfrac{1\! 1}2}}}} \ .  
\ee
As in the other cases the non-degenerate orbit provides the homogeneous solution to the differential equations of~\cite{Bossard:2015uga} and it was shown in particular in~\cite{Bossard:2015oxa} that the above Eisenstein series encodes all the relevant information about supergravity divergences up to three loops.

\subsection{$D<3$}

We can also treat the expression~\eqref{eq:a4} formally in dimensions when the hidden symmetry is thought to be of Kac--Moody type~\cite{Julia,West:2001as,Damour:2002cu}. A full exceptional field theory has not been developed in these cases. For $D=2$ one can define the four-scalar amplitude in supergravity and the two-derivative effective theory is known to admit a Kac--Moody $E_9$ symmetry \cite{Nicolai:1987kz}. A closed algebra of generalised diffeomorphisms for the corresponding exceptional field theory has been defined in \cite{Bossard:2017aae}, and it involves exceptional coordinates in the expected highest weight module $\overline{\bf R}_{\alpha_d}$. As we explain in Appendix~\ref{app:E9}, the situation is then essentially as much in control as for $D=3$, so that one arrives to the same formula \eqref{eq:a4}. For $d\ge 8$, this formula indicates that one can neglect the lower rank contribution in the limit $\mu\rightarrow 0$, so that one only gets the maximal rank contribution with a finite coefficient. In this section we shall also extrapolate these formulas for $d>9$, although there is no clear scattering amplitude defining the coupling in this case.

From the analysis of the $D=3$ case above, we anticipate that there should only be a `fundamental' series on the last node of the $E_d$ Dynkin diagram in Figure~\ref{fig:dynk}. Using the properties of Kac--Moody Eisenstein series~\cite{Garland1,Garland2,Fleig:2012xa,Carbone:2017} we can address this question. As we discuss in more detail in Appendix~\ref{app:E9} one can relate the constrained Epstein sums over 1/2-BPS charges to Langlands Eisenstein series on the completed Kac--Moody group that are also discussed in the appendix.

For $D=2$ ($d=9$) we are in the affine $E_9$ case. Due to the degenerate Cartan matrix one has to treat the derivation (that is used to desingularise the Cartan matrix) separately~\cite{Garland1,Garland2}. There is a parameter $v$ associated with the derivation direction in the affine Lie algebra that is dual to the null root. We discuss more details related to this subtlety in Appendix~\ref{app:E9}. One has that the fundamental series for $E_9$ satisfies
\begin{align}
\label{eq:affFR1}
E_{s\Lambda_{9}} =  \frac{\xi(2s-13)\xi(2s-14)\xi(2s-15)}{\xi(2s)\xi(2s-6)\xi(2s-10)} v^{14-4s}E_{(s-6)\Lambda_3+(8-s)\Lambda_7}\,,
\end{align}
where we use the notational conventions discussed around~\eqref{eq:ESconv}.
For $s=6$ one deduces then that the non-degenerate orbit contribution in~\eqref{eq:a4} is
\begin{align}
40\xi(2)\xi(3)\xi(4) v^2 E_{\alpha_7,2} = 40 \xi(2)\xi(6) \xi(12) v^{12} E_{\alpha_{9},6}
\end{align}
and this agrees with the one-loop result of~\cite{Bossard:2015foa}, see also Appendix~\ref{app:E9}. 

As stressed in the introduction the cases $E_{10}$ and $E_{11}$ are more formal, but we will now show that they work in an exactly parallel manner. 

For $D=1$ ($d=10$)  the conjectured symmetry group is the hyperbolic $E_{10}$ and we have formally that an Eisenstein series on the fundamental node satisfies the functional relation
\begin{align}
E_{s\Lambda_{10}} =  \frac{\xi(2s-15)\xi(2s-16)\xi(2s-17)}{\xi(2s)\xi(2s-7)\xi(2s-11)} E_{(s-\tfrac{13}2)\Lambda_3+(9-s)\Lambda_8}
\end{align}
so that the non-degenerate orbit in~\eqref{eq:a4} is related to a fundamental series by
\begin{align}
40\xi(3)\xi(4)\xi(5) E_{\alpha_8,5/2} = 40 \xi(2)\xi(6)\xi(13)  E_{\alpha_{10},13/2} = \cE_{\gra{0}{1}}^{\ord{\scalebox{0.6}{1-loop}}}\,,
\end{align}
where the last step shows that this is formally equal to the contribution from the one-loop exceptional field theory amplitude in~\cite[Eq.~(3.9)]{Bossard:2015foa}.

For $D=0$ ($d=11$) the conjectured symmetry group is the Lorentzian $E_{11}$. The functional relation in this case reads 
\begin{align}
E_{s\Lambda_{11}} =  \frac{\xi(2s-17)\xi(2s-18)\xi(2s-19)}{\xi(2s)\xi(2s-8)\xi(2s-12)} E_{(s-7)\Lambda_3+(10-s)\Lambda_{9}}\,.
\end{align}
At $s=7$ this gives the desired relation between the non-degenerate orbit in~\eqref{eq:a4} and the one-loop calculation:
\begin{align}
40\xi(4)\xi(5)\xi(6) E_{\alpha_9,3} = 40 \xi(2)\xi(6)\xi(14)  E_{\alpha_{11},7} = \cE_{\gra{0}{1}}^{\ord{\scalebox{0.6}{1-loop}}}\,.
\end{align}

\section{One-quarter BPS contributions}
\label{sec:14BPS}

One way to extract the 1/4 BPS states contribution is to consider the superstring amplitude. $E_d(\mathds{Z})$ relates all 1/2 BPS states to 11-dimensional supergravity torus Kaluza--Klein states. Similarly,  $E_d(\mathds{Z})$ relates all 1/4 BPS states to perturbative string theory states with torus winding and momenta that are not orthogonal (do not satisfy level matching). The 1-loop string theory contribution to the $\nabla^4 R^4$ coupling is given by an integral of a modular graph function against the Narain theta function $\Gamma_{d-1,d-1}$ associated with the torus $T^{d-1}$. The modular graph function in this case is well-known to be proportional to the real analytic Eisenstein series $E_2$~\cite{Obers:1999um,Green:2010wi,Angelantonj:2012gw} and this leads to
\be 
 \label{String1Loop} 
\cE_{\gra{1}{0}}^{\scalebox{0.7}{String (1-loop)}} = 4 \pi  g_s^{-2\frac{d+1}{9-d}} \int_{\cF_1} \frac{d^2 \tau}{\tau_2^{\; 2}} \xi(4) E_2(\tau) \tau_2^{\; \frac{d-1}{2}} \sum_{Q\in \mathds{Z}^{d-1,d-1}} e^{-\pi \tau_2 g(Q,Q) + i \pi \tau_1 \langle Q,Q\rangle} \ .
\ee
$\mathcal{F}_1$ denotes the fundamental domain of the inequivalent toroidal world-sheets parametrised by the world-sheet modulus $\tau$. $Q$ denotes the momentum and winding charges of the string on $T^{d-1}$.

The standard unfolding procedure, including an appropriate regulator~\cite{Zagier:1981,Angelantonj:2011br}, permits to compute the integral~\eqref{String1Loop} and to recover the $O(d-1,d-1)$ vector Eisenstein series of weight $\frac{d+1}{2}$. This computation suggests that only 1/2 BPS states contribute to the amplitude, because the integral over $\tau_1$ then enforces the level matching condition $\langle Q,Q\rangle= 0$ satisfied by 1/2 BPS string states. However, we know that 1/4 BPS states contribute as well when $d>0$~\cite{Green:1999pv,Green:2011vz}. To recover the complete set of states contributing to the amplitude from~\eqref{String1Loop} we substitute the formal identity 
\be 
1 = \lim_{\epsilon\rightarrow 0} \sum_{\gamma \in \mathds{Z} \backslash  PSL(2,\mathds{Z})} \tau_{2,\gamma}^{\; \epsilon} 
\ee
by analytic continuation of the function for Re$(\epsilon) >1$. As a consequence, we are saying that \eqref{String1Loop} is formally equal to the same integral over a complete unit strip in the upper complex half-plane $\cH^+$ when we freely unfold this coset sum and then take the limit again. In the integral over $\cH^+$ we can then substitute the Fourier expansion of $E_2(\tau)$ and obtain for the non-zero Fourier coefficients\footnote{\label{fn:EsFourier}The general expansion of $E_s(\tau)$ for $\tau=\tau_1+i \tau_2$ is
\begin{align}
\label{eq:EsFourier}
E_s(\tau) = \sum_{\gamma\in \mathds{Z}\backslash SL(2,\mathds{Z})} \!\!\! \left( \mathrm{Im}\, \tau_\gamma\right)^s = \tau_2^s + \frac{\xi(2s-1)}{\xi(2s)} \tau_2^{1-s} + \frac{2}{\xi(2s)} \tau_2^{1/2} \sum_{n\neq 0}  |n|^{s-1/2} \sigma_{1-2s}(n) K_{s-1/2}(2\pi |n| \tau_2) e^{2\pi i n \tau_1}\,.\nn
\end{align}
Here, $\sigma_k(n)=\sum_{d|n} d^k$ is the divisor sum of $n$; the variable $d$ runs over the positive divisors of $n$.}
\bea 
\label{eq:FI}
&& 8 \pi  g_s^{-2\frac{d+1}{9-d}} \sum_{n\neq 0} \sum_{Q \in\mathds{Z}^{d-1,d-1}}\int_{\mathds{Z}\backslash \cH^+} \frac{d\tau}{\tau_2^2}  \tau_2^{d/2} \sigma_{3}(n) |n|^{-3/2} K_{\frac32}(2\pi|n|\tau_2) e^{2\pi i n \tau_1 - \pi i \tau_1 \langle Q , Q\rangle-\pi \tau_2 g(Q,Q)}\nn\\
&=& 8\pi g_s^{-2\frac{d+1}{9-d}} \int_{0}^\infty \frac{d\tau_2}{\tau_2} \sum_{\substack{Q\in \mathds{Z}^{d-1,d-1}\\ \langle Q,Q\rangle \ne 0}} \frac{\sigma_3(|\frac{\langle Q,Q\rangle}{2}|)}{|\frac{\langle Q,Q\rangle}{2}|^{\frac32} } K_{\frac32}(2\pi \tau_2  |\tfrac{\langle Q,Q\rangle}{2}|) \tau_2^{\frac{d-2}{2}}e^{-\pi \tau_2 g(Q,Q)} \CR
&=& 2 \pi^{\frac{5-d}{2}} \Gamma(\tfrac{d-5}{2}) \sum_{\substack{Q\in \mathds{Z}^{d-1,d-1}\\ \langle Q,Q\rangle \ne 0}} \sigma_3(|\tfrac{\langle Q,Q\rangle}{2}|)   \frac{g_s^{\frac{4}{9-d}}  g(Q,Q) + (d-3) g_s^{\frac{4}{9-d}} |\frac{\langle Q,Q\rangle}{2}|}{(g_s^{\frac{4}{9-d}} |\frac{\langle Q,Q\rangle}{2}|)^3(g_s^{\frac{4}{9-d}}  g(Q,Q) + 2 g_s^{\frac{4}{9-d}} |\frac{\langle Q,Q\rangle}{2}|)^{\frac{d-3}{2}}} \ .  
\eea
The integral over the Bessel function can be carried out for example using the exact asymptotic expansion around $\tau_2\to \infty$: $K_{\frac32} (x) = \sqrt{\frac{\pi}{2x}} e^{-x} (1 + \tfrac1x)$. Note that the integral over $\tau_1$ has produced the constraint $\langle Q, Q\rangle=2n \neq 0$ corresponding to 1/4 BPS states. 

The quantities involving $Q$ appearing in the last expression in~\eqref{eq:FI} can be reinterpreted in terms of specific  representatives of charges $\Gamma$ in U-duality multiplets of 1/4 BPS states as follows. For all 1/4 BPS charges $\Gamma$ there is an element $\gamma \in P_1\backslash E_d$ such that $\Gamma$ lies in the vector representation highest degree component of the corresponding decomposition of the representation $R(\alpha_d)$ of $E_{d(d)}$ under $O(d-1,d-1)$. We denote this component in $\mathds{Z}^{d-1,d-1}$ by $Q$ and have
\be 
\label{ZtopLR}
|Z(\Gamma)|^2 = g_s^{\frac{4}{9-d}} g(Q,Q)\ ,\quad \Delta(\Gamma)= g_s^{\frac{8}{9-d}}\, \bigl| \tfrac{\langle Q,Q\rangle}{2} \bigr|^2 \ . \ee
Taking the Poincar\'e sum of this contribution under the \textit{full} $E_{d}(\mathds{Z})$, one obtains the manifestly U-duality invariant $\nabla^4 R^4$ threshold function corresponding to 1/4 BPS states from the T-duality invariant string expression~\eqref{eq:FI} as
\be
\label{eq:14BPSU}
\cE_{\gra{1}{0}}^{\scalebox{0.6}{(1-loop) 1/4-BPS}}  = 2 \frac{\xi(d-5)}{\zeta(d-5)}\sum_{\substack{\Gamma \in \mathds{Z}^{d(\alpha_d)} \\\Gamma \times \Gamma \ne 0\\ I_4^\prime(\Gamma) = 0}}  \frac{\sigma_{3}(\Gamma \times \Gamma) }{\Delta(\Gamma)^{\frac32} }  \frac{|Z(\Gamma)|^2  + (d-3) \sqrt{ \Delta(\Gamma)} }{( |Z(\Gamma)|^2 + 2  \sqrt{ \Delta(\Gamma)} )^{\frac{d-3}{2}}} \ .  
\ee
The constraint on the sum is exactly the 1/4 BPS constraint~\cite{Obers:1998fb}. The superscript `1-loop' for this U-duality invariant function refers to the loop order from the point of view of the effective field theory analysis. 

The amplitude~\eqref{eq:14BPSU} is indeed consistent with the expected result: The factor of $( |Z(\Gamma)|^2 + 2  \sqrt{ \Delta(\Gamma)} )^{\frac{d-3}{2}}$ is the 1/4-BPS mass to the power corresponding to a 1-loop box diagram in $11-d$ dimensions and  $ \sigma_3(\Gamma \times \Gamma) $ is the twelfth helicity supertrace of the 1/4 BPS states computed in perturbative string theory~\cite{Kiritsis:1997hj,Bossard:2016hgy}. The numerator is determined by supersymmetry such that the function $ \scalebox{0.7}{$\frac{1}{\Delta(\Gamma)^{\frac32} }  \frac{|Z(\Gamma)|^2  + (d-3) \sqrt{ \Delta(\Gamma)} }{( |Z(\Gamma)|^2 + 2  \sqrt{ \Delta(\Gamma)} )^{\frac{d-3}{2}}}$} $ satisfies the differential equations imposed by supersymmetry on $\cE_{\gra{1}{0}}^{\scalebox{0.6}{(1-loop) 1/4-BPS}}$. In particular, the full $\cE_{\gra{1}{0}}^{\ord{D}}$ must satisfy a Laplace equation of the form ($D=11-d$ as always)
\begin{align}
\Bigl(  \Delta_{E_d} - 5 \frac{(d+1)(4-d)}{9-d}  \Bigr) \cE_{\gra{1}{0}}^{\ord{D}} = 40\zeta(2) \delta_{d,4} + 7\cE_{\gra{0}{0}}^{\ord{6}}\delta_{d,5} 
\end{align}
that is homogeneous away from $D=6$ and $D=7$.  Considering the `theta kernel' function appearing in~\eqref{String1Loop} one finds indeed that\footnote{For which $\Delta_{E_d} = \frac{(9-d)(31-3d)}{32} g_s \frac{\partial\; }{\partial g_s} g_s \frac{\partial\; }{\partial g_s} -\frac{42+d(d-27)}{8}  g_s \frac{\partial\; }{\partial g_s} + \frac{1}{2} ( g(Q,Q)^2 - \langle Q,Q\rangle^2 )  \frac{\partial^2\; }{\partial g(Q,Q)^2}+ \frac{d-1}{2}  \frac{\partial\; }{\partial g(Q,Q)}$.}
\bea  
&& \Bigl(  \Delta_{E_d} - 5 \frac{(d+1)(4-d)}{9-d}  \Bigr) \Bigl( g_s^{-2\frac{d+1}{9-d}} \tau_2^{\; \frac{d-1}{2}} e^{-\pi \tau_2 g(Q,Q) + i \pi \tau_1 \langle Q,Q\rangle} \Bigr) \CR
&=& \frac{1}{2} ( \Delta_\tau - 2) \Bigl( g_s^{-2\frac{d+1}{9-d}}  \tau_2^{\; \frac{d-1}{2}} e^{-\pi \tau_2 g(Q,Q) + i \pi \tau_1 \langle Q,Q\rangle}\Bigr)  \ .
\eea
This implies that the function that this kernel is integrated against in a `theta lift' must be an eigenfunction of the upper complex half plane Laplacian $\Delta_\tau$ of eigenvalue $2$. The same construction can be used to show that all the Casimir differential operators on $E_{d(d)}/ K(E_d)$ take the correct eigenvalues imposed by supersymmetry, provided that the source function satisfies this Laplace equation. The only two solutions associated to 1/4 BPS charges with a non-trivial dependence in $ e^{2\pi i \tau_1 n}$ with $n\ne 0$ are $e^{\pm 2\pi |n| \tau_2+2\pi i n \tau_1 }( 1 - \frac{1}{2\pi |n| \tau_2})$, where the minus sign occurs for the Fourier coefficient of $E_2(\tau)$ and the other, exponentially growing, solution is normally eliminated by moderate growth conditions.  This second solution gives a function of the form $ \scalebox{0.7}{$\frac{1}{\Delta(\Gamma)^{\frac32} }  \frac{|Z(\Gamma)|^2  - (d-3) \sqrt{ \Delta(\Gamma)} }{( |Z(\Gamma)|^2 - 2  \sqrt{ \Delta(\Gamma)} )^{\frac{d-3}{2}}}$} $ that is singular at finite values of the moduli and does not reproduce the expected mass term for a 1/4 BPS state contribution. We have checked in \cite{Bossard:2015foa} that these are  indeed the two unique solutions to the tensorial equations imposed by supersymmetry for $d=4$. 

To evaluate~\eqref{eq:14BPSU}, we go back to its form before integration written as the $E_d(\mathds{Z})$ Poincar\'e sum of the perturbative string theory contribution. For this purpose we note that for any 1/4 BPS charge, there exists an $E_{d}(\mathds{Z})$ element to bring the cross product $\Gamma\times \Gamma$ into the highest weight component of the highest weight representation $\Lambda_1$ of stabiliser $P_1(\mathds{Z}) \subset E_{d}(\mathds{Z})$. This is precisely the decomposition in which the charge $\Gamma$ is represented by the highest weight component vector $Q \in \mathds{Z}^{d-1,d-1}$ and $\Gamma\times \Gamma$ is $ \tfrac{\langle Q,Q\rangle}{2}$. We conclude that we have a representation of the 1/4 BPS threshold function as a theta lift of an $E_2(\tau)$ Fourier mode 
\be 
\cE_{\gra{1}{0}}^{\scalebox{0.6}{(1-loop) 1/4-BPS}}  = 8 \pi \sum_{\gamma \in P_1\backslash E_d}  g_{s, \gamma}^{\; -2\frac{d+1}{9-d}} \sum_{n\ne 0} \frac{\sigma_{3}(|n|)}{n^{2} }    \int_{\mathds{Z}\backslash \cH^+} \frac{d^2 \tau}{\tau_2^{\; 2}} \sqrt{|n|\tau_2} K_{\frac32}(2\pi |n| \tau_2) e^{2\pi i n \tau_1 } \Gamma_{d-1,d-1}(\tau,g_\gamma)  
\ee
with the Narain genus-one partition function 
\be   
\Gamma_{d-1,d-1}(\tau,g)  = \tau_2^{\; \frac{d-1}{2}}  \sum_{Q\in \mathds{Z}^{d-1,d-1}} e^{-\pi \tau_2 g(Q,Q) + \pi i \tau_1 \langle Q,Q\rangle}\ . 
\ee
We shall formally now fold this integral by defining the Eisenstein series from the Poincar\'e sum of the Whittaker function 
\be  
\label{eq:WhittakerPoincare}
\sum_{\gamma \in P_1\backslash  PSL(2,\mathds{Z})}  \sqrt{|n|\tau_{2}} K_{s-\frac12}(2\pi |n| \tau_{2}) e^{2\pi i n \tau_{1} } \Big|_\gamma  = \frac{\pi}{(2s-1) \cos( \pi s)\xi(2s-1)} \frac{\sigma_{2s-1}(|n|)}{|n|^{s-1}} E_{s}(\tau) \ . 
\ee
Formally, the left-hand side defines an $SL(2,\mathds{Z})$ invariant function with eigenvalue $s(s-1)$ under the Laplacian and thus should be proportional to the Eisenstein series $E_s(\tau)$ for real $s$. The above equation provides the proportionality factor in a formal way but the actual Poincar\'e sum does not converge for any $s$. It can be written as the difference of two Niebur--Poincar\'e series that are absolutely convergent on two different domains (Re$(s)>1$ and Re$(1-s)>1$ respectively), see \cite{Angelantonj:2012gw}
\be 
\sum_{\gamma \in P_1 \backslash  PSL(2,\mathds{Z})}  \sqrt{|n|\tau_{2}} K_{s-\frac12}(2\pi |n| \tau_{2}) e^{-2\pi i n \tau_{1} } \Big|_\gamma  = \frac{\sqrt{\pi}}{2 \cos( \pi s)} \Bigl( \frac{ \cF(s,n,0)}{4^s \Gamma(s+\frac{1}{2})} -  \frac{ \cF(1-s,n,0)}{4^{1-s} \Gamma(\frac{3}{2}-s)} \Bigr) \ . 
\ee
The limit $s\to 2$ is nevertheless regular and one obtains from~\eqref{eq:WhittakerPoincare} that 
\be 
\cE_{\gra{1}{0}}^{\scalebox{0.6}{(1-loop) 1/4-BPS}}  = 480 \frac{\xi(4)}{\xi(3)} \sum_{\gamma \in P_1\backslash E_d} g_{s, \gamma}^{\; -2\frac{d+1}{9-d}} \sum_{n> 0} \frac{\sigma^{\; 2}_3(n)}{n^3}     \int_{\cF_1} \frac{d^2 \tau}{\tau_2^{\; 2}} E_2(\tau) \Gamma_{d-1,d-1}(\tau,g_\gamma)\ , 
\ee
where the Fourier sum is now only over $n>0$. The sum $ \sum_{n> 0} \frac{\sigma^{\; 2}_3(n)}{n^3}$ still diverges, but using a zeta function regularisation via the Ramanujan identity (see Appendix~\ref{app:RI})\footnote{The dimensional regularisation gives naturally $E_{2+\epsilon}(\tau)$ and the sum $\sum_{n> 0} \frac{\sigma^{\; 2}_{3+2\epsilon}(n)}{n^{3+2\epsilon}} = \zeta(3+2\epsilon)\zeta(-3-2\epsilon) \zeta(0) $ does not converge either.}
\be  
\label{eq:RI}
\sum_{n>0} \frac{\sigma_3^2(n)}{n^3} \to -\frac{\zeta(3)}{240}
\ee
one obtains that
\begin{align} 
\cE_{\gra{1}{0}}^{\scalebox{0.6}{(1-loop) 1/4-BPS}}  &= -4\pi  \xi(4) \sum_{\gamma \in P_1\backslash E_d} g_{s, \gamma}^{\; -2\frac{d+1}{9-d}}    \int_{\cF_1} \frac{d^2 \tau}{\tau_2^{\; 2}} E_2(\tau) \Gamma_{d-1,d-1}(\tau,g_\gamma) \nn\\
& = -8 \pi \xi(4) \xi(d+1) \sum_{\gamma\in P_1\backslash E_d}  g_{s, \gamma}^{\; -2\frac{d+1}{9-d}}   E^{SO(d-1,d-1)}_{V,\frac{d+1}{2}}(g_\gamma)\nn\\
&= - 8\pi \xi(4)\xi(d+1) E_{\alpha_d,\frac{d+1}{2}}
\end{align}
where we have first carried out the (regularised) theta lift using for instance the results of~\cite{Angelantonj:2011br} to obtain a vector Eisenstein series on the T-duality group $SO(d-1,d-1)$\footnote{In our notation, the general formula is
$
\int\limits_{\mathcal{F}_1} \frac{d^2\tau}{\tau_2^2} E_s(\tau) \Gamma_{d-1,d-1}(\tau) = 2\xi(2s+d-3) E^{SO(d-1,d-1)}_{V,s+\frac{d-3}{2}}
$.} 
and then performed the Poincar\'e sum over the U-duality group starting from the constant term in the expansion of $E_{\alpha_d,\frac{d+1}{2}}$ along the T-duality subgroup.

Recalling that $\cE_{\gra{1}{0}}^{\scalebox{0.6}{(1-loop) 1/2-BPS}} = 8\pi \xi(4)\xi(d+1) E_{\alpha_d,\frac{d+1}{2}}$ from~\cite{Bossard:2015foa}, this formally proves a claim that was made there: The 1/4 BPS state contribution cancels precisely the divergent 1/2 BPS amplitude, such that the contributions of all states with $\gcd( \Gamma\times \Gamma) = n$ give the same contribution with a weight that gives a divergent overall factor 
\be  
\label{eq:cancel14}
\cE_{\gra{1}{0}}^{\scalebox{0.6}{(1-loop) 1/2-BPS+1/4-BPS}}  =8 \pi  \Bigl( 1 + \frac{240}{\zeta(3)} \sum_{n>0} \frac{\sigma^{\; 2}_{3}(n)}{n^{3}} \Bigr)\xi(4)\xi(d+1) E_{\alpha_d, \frac{d+1}{2}} \  \ , 
 \ee
which formally vanishes in zeta regularisation.

Although we have been manipulating several expressions formally in this section in order to regularise infinite sums without defining a proper analytic continuation from absolutely convergent sums, it seems reasonable to assume that there could be proved using well-defined analytic continuation in an appropriate regularisation scheme. Assuming this is the case, one would conclude that the $\nabla^4 R^4$ threshold function $\cE_{\gra{1}{0}}$ comes entirely from the exceptional field theory 2-loop 1/2 BPS contribution. Note that the same argument at string two loops exhibits that there is no contribution from 1/4 BPS states to the $\nabla^4 R^4$ threshold function since the genus 2 integrand is just the Narain partition function. 

Let us note finally that the computation above can be interpreted in perturbative string theory also as follows. The function $E_2(\tau)$ appearing in the theta lift~\eqref{String1Loop} has three distinct pieces in its Fourier expansion, namely two constant terms and the non-zero Fourier modes, see footnote~\ref{fn:EsFourier}. The two constant terms give twice the contribution in $8\pi \xi(4)\xi(d+1) E^{SO(d-1,d-1)}_{V,(d+1)/2} $ (using the Langlands functional relation for vector series for the second), whereas the $1/4$-BPS sum over strings with non-orthogonal winding and momenta gives formally the same contribution with a minus sign (cf.~\eqref{eq:cancel14}), to eventually reproduce the correct perturbative contribution. If one were to compute the one-loop amplitude within an effective field theory with all massive states in string theory, this is the infinite sum one would need to regularise. Modular invariance of string perturbation theory permits to combine all these states in a manifestly finite form, which regularises the infinitely many Feynman diagrams one would find instead in field theory. Such infinite sums are therefore also expected in the non-perturbative effective theory. One may expect that a consistent formulation of M-theory would provide the appropriate integral regularising this sum, with an appropriate notion of modular invariance.

\section{Comments on systematics of BPS corrections}
\label{sec:system}

We have seen above in~\eqref{eq:cancel14} that the 1/4 BPS contribution that follows from the perturbative string theory one-loop calculation and completed to a U-duality invariant $\nabla^4R^4$ threshold function as in~\eqref{eq:14BPSU} cancels formally the one-loop 1/2 BPS contribution computed in exceptional field theory in~\cite{Bossard:2015foa}.  It follows that the entire contribution to the $\nabla^4 R^4$ coupling in the loop expansion involving all BPS states appears at 2-loop. In this section we shall discuss the analogous structures that one expects for the $\nabla^6 R^4$ coupling, and how they are compatible with the result of the 3-loop computation we have carried out in this paper. 

Before embarking on this discussion we want to clarify the distinction between BPS solitons and BPS instantons, and their respective `non-renormalisation theorems'. The BPS solitons that contribute to the low-energy effective action are the BPS black hole solutions in supergravity. In type II string theory, they correspond to fundamental strings, D$p$-branes, NS5-branes and KK-branes that wrap the $T^{d-1}$ torus such that they are effectively point-like particles in the uncompactified $D$ dimensions. The BPS instantons depend on the perturbative frame. In type II string theory they are the Euclidean D$p$-branes and NS5-brane that wrap the $T^{d-1}$ torus along all their directions. In $D=11$ supergravity they can be defined as M-theory instantons described by Euclidean M2- or M5-branes wrapping the $T^d$ torus.\footnote{In $D=3$ one has moreover Kaluza--Klein instantons that contribute, but the general interpretation of the various contributions is less clear in this case since the solitons do not have a charge in the discrete lattice in three dimensions, but the scalar fields instead admit a non-trivial monodromy in $E_d(\mathds{Z})\backslash E_d(\mathds{R}) / K(E_d)$.} One can also formally consider the large radius limit as a perturbative theory in the inverse radius, in which case the instantons can sometimes be identified as black holes in $D+1$ dimensions compactified over the thermal time circle \cite{Green:2011vz,Gunaydin:2005mx,Bossard:2016zdx,Bossard:2016hgy,Bossard:2017wum}. 

The set of instantons that can contribute to a higher-derivative coupling is mathematically equivalent to the so-called wave-front set of the corresponding automorphic function of the U-duality group~\cite{Pioline:2010kb,Green:2011vz,Fleig:2015vky}. The wave-front set in mathematical terms is a description of all non-vanishing Fourier coefficients an automorphic function or form has. More precisely, an automorphic form belongs to an automorphic representation, and the wave-front set is attached to the automorphic representation. One says that an automorphic representation is small, if most of the Fourier coefficients vanish. For a given representation, the wave-front set is the closure of typically a single nilpotent orbit of the hidden symmetry group $E_{d}(\mathds{C})$ in the Zariski topology.\footnote{Below we will label the nilpotent orbits by their Bala--Carter type~\cite{CollingwoodMcGovern,Spaltenstein}. Type $A_1$ describes the minimal nilpotent orbits, type $2A_1$ the next-to-minimal and the orbits can be arranged on a Hasse diagram. The minimal orbit characterises a unique automorphic representation for $d\ge 5$ and the next-to-minimal orbit characterises a unique automorphic representation for $d\ge 7$.} As instanton corrections are associated with non-trivial Fourier coefficients of the automorphic threshold function~\cite{Green:1997tv,Obers:1999um,Green:2010wi,Pioline:2010kb,Green:2011vz,Fleig:2015vky}, the wave-front set encodes which types of instantons contribute to a given threshold function $\cE_{\gra{p}{q}}^D$. The wave-front set can equivalently be defined from a set of differential equations satisfied by the automorphic forms belonging to a given representation~\cite{Borho:1982,Barbasch:1985}. For the first few couplings in the low-energy expansion, $R^4$, $\nabla^4 R^4$ and $\nabla^6 R^4$, supersymmetry constrains the couplings to satisfy differential equations, which imply that the only instanton corrections contributing to them are at most  respectively 1/2 BPS, 1/4 BPS, or 1/8 BPS, respectively \cite{Bossard:2014lra,Bossard:2014aea,Bossard:2015uga}. The differential equations following from the supersymmetry Ward identities, and the property that only certain supersymmetric instantons can possibly contribute to a protected coupling, are two aspects of the same mathematical concept, the wave-front set discussed above. More precisely, there are two types of $\nabla^6 R^4$ supersymmetry invariants in dimensions $4\le D\le 7$ \cite{Bossard:2015uga}, one that we shall call \textit{chiral} and that satisfies a homogeneous differential equation and a second one that we shall refer to as \textit{non-chiral}  and that satisfies an inhomogeneous differential equation. The chiral invariant is associated to an automorphic representation of Bala--Carter type $2A_1$ for $D\ge 5$, {\it i.e.} 1/4 BPS, and of Bala--Carter type $3A_1$ for $D=4$, {\it i.e.} 1/8 BPS (chiral), while the homogeneous solution to the equation of the non-chiral invariant is associated to a Bala--Carter type $A_2$ automorphic representation,   {\it i.e.} 1/8 BPS.\footnote{One can often think of the Bala--Carter type $n A_1$ as being associated to a multiple intersection of $n$ orthogonal 1/2 BPS instantons, which therefore preserves  $1/2^n$ of the supersymmetry, and the Bala--Carter type $A_2$ as being associated to special intersections of 1/2 BPS instantons, as for a D0-D6 type IIA bound state~\cite{Argurio:1998cp,Witten:2000mf,Denef:2007vg}, which preserves  $1/8$ of the supersymmetry.} The unique automorphic form satisfying the corresponding homogeneous differential equation of the chiral invariant is the `fundamental' Eisenstein series $E_{\alpha_d,\frac{d+3}{2}}$ of Bala--Carter type $2A_1$ (or $3A_1$ for $d=7$) while the unique homogeneous solution of the inhomogeneous equation for the non-chiral invariant is the Eisenstein series associated to the adjoint representation at a particular value of the weight parameter ({\it i.e.} $E^{E_7}_{\alpha_1,6}$, $E^{E_6}_{\alpha_2, 9/2}$, $E^{D_5}_{\alpha_2,7/2}$, $E^{A_4}_{[3\, 0\, 0\,5/2]}$)\footnote{Here, we have used the Bourbaki labelling of the algebras $D_5$ and $A_4$. If one used instead the induces `exceptional' labelling of $E_5$ and $E_4$ that comes from diagram~\ref{fig:dynk}, the functions would be $E^{E_5}_{\alpha_3,7/2}$ and $E^{E_4}_{[3\, 5/2\, 0\,0]}$.} of Bala--Carter type $A_2$. The full inhomogeneous solution can be constructed formally (up to regularisation issues discussed below) from the particular solution provided by the two-loop exceptional field theory calculation in $D>3$~\cite{Bossard:2015foa}, {\it i.e.} it is determined by the 2-loop 1/2 BPS states contribution. The quadratic source term in the inhomogeneous equation of the 1/8 BPS coupling has as maximal orbit in the wave-front set the orbit of Bala--Carter type $A_2$, so it is natural to consider the full inhomogeneous non-chiral solution to be characterised by this wave-front set \cite{Bossard:2015oxa} even if it does not belong to an automorphic representation in the strict mathematical sense. The chiral contribution to the coupling has the maximal orbit $3A_1$ in four dimensions, so that the full $\nabla^6R^4$ coupling wave-front set then has two maximal orbits.

There is no clear non-renormalisation theorem for the BPS solitons, but one can get some insights from string perturbation theory. The type of BPS states that can contribute to a given coupling in string perturbation theory does not  depend only on the type of coupling but also on the loop order. At genus $1\leq g \leq 3$, the $g$-loop contribution to such couplings is defined in perturbation theory as the theta lift of a particular automorphic form of $Sp(2g)$  with the genus $g$ Narain theta function. For the $R^4$ coupling at 1-loop, the $\nabla^4 R^4$ coupling at 2-loops, and the $\nabla^6 R^4$ coupling at 3-loops, the automorphic form is a constant, so that the only states that contribute in the loop satisfy the level matching condition and are 1/2 BPS string states.  Since perturbative 1/4 BPS states do not contribute in this case, one concludes using U-duality that the same is true for non-perturbative 1/4 BPS states, so that these couplings only receive corrections from 1/2 BPS solitons at these loop orders. A similar argument using U-duality and the string amplitude cannot be applied for the 1/8 BPS states, but it is legitimate to assume that the absence of 1/4 BPS corrections implies the absence of 1/8 BPS corrections at the same order. The mechanisms responsible for the cancellation of contributions coming from BPS multiplets not preserving enough supersymmetry are indeed always ordered, because they are usually associated with the matching of a certain number of fermion zero modes with the number of supercharges that annihilate an operator.

At one loop one can identify the perturbative 1/4 BPS states contributions to the $\nabla^4 R^4$ coupling as we did in the last section. The same argument can be applied to the $\nabla^6 R^4$ coupling where we recall that the exceptional field theory one-loop calculation gives the 1/2 BPS contribution proportional to the (fundamental) Eisenstein series $E_{\alpha_d,\frac{d+3}{2}}$~\cite{Bossard:2015foa}. Repeating for $\nabla^6 R^4$ the same steps as performed in Section~\ref{sec:14BPS} for $\nabla^4R^4$, one extracts from the string theory one-loop amplitude the 1/4 BPS contribution by U-duality completion, such that 
\be
\cE_{\gra{0}{1}}^{\scalebox{0.6}{(1-loop) 1/2-BPS+1/4-BPS}}  =40 \Bigl( 1 - \frac{504}{\zeta(5)} \sum_{n>0} \frac{\sigma^{\; 2}_{5}(n)}{n^{5}} \Bigr)\xi(2)\xi(6)\xi(d+3) E_{\alpha_d, \frac{d+3}{2}} = 0 \  ,  
 \ee
Again, these two contributions cancel formally using zeta regularisation. Similar to~\eqref{eq:14BPSU}, one can relate the divisor sum to helicity supertraces $B_n$ of 1/4 BPS states, where  \cite{Kiritsis:1997hj}
\be 
B_n(\Gamma)  =\frac{(-1)^{\frac{n}{2}}}{n!}  \mbox{Tr}^\prime_\Gamma\, (-1)^{2J_3} ( 2 J_3)^n \ ,
\ee
is the supertrace over the space of states with charge $\Gamma$ and the prime indicates that the bosonic zero mode corresponding to the center of mass has been removed.  The multiplicity of the 1/4 BPS contributions to  the $\nabla^6 R^4$ above is then the contribution  of 1/4 BPS multiplets of charge $\Gamma$ to the helicity supertrace $6 B_{14}+2 B_{12}=\sigma_5( \Gamma \times \Gamma)$. It is rather natural that the contribution to these couplings is related to the helicity supertrace, because they are the unique observables that preserve these precise fractions of supersymmetry. Extrapolating this structure to the 1/8 BPS solitons contribution in $D=4$ and $5$ (there are no 1/8 BPS black hole solutions in $D\ge 6$), one expects that they should not contribute to the $\nabla^4 R^4$ coupling, and that the contribution to the $\nabla^6 R^4$ coupling of 1/8 BPS states of charge $\Gamma$ should be proportional to its contribution to the helicity supertrace $B_{14}$, that is proportional to the Fourier coefficient of the Jacobi  function $- \vartheta_1(z,\tau)^2/ \eta(\tau)^6$ in $D=4$ \cite{Shih:2005qf}.

Extrapolating this structure to 2-loops, and consistently with the non-renormalisation theorem for the instanton corrections discussed above, one concludes that the $R^4$ coupling receives perturbative corrections only from 1/2 BPS states at 1-loop,  the $\nabla^4 R^4$ coupling receives corrections from 1/4 BPS states only at 1-loop and 1/2 BPS states up to 2-loop, and the $\nabla^6 R^4$ coupling receives 1/8 BPS states contributions only at 1-loop, 1/4 BPS states contributions up to 2-loop, and 1/2 BPS states contributions up to 3-loop. Summarising the possible contributions to $\nabla^6R^4$, including the ones we have already computed, one obtains the table 
\begin{center}
\begin{tabular}{c|ccc}
$\nabla^6 R^4$ & $\frac12$ BPS & $\frac14$ BPS & $\frac18$ BPS\\[2mm]\hline\\[-2mm]
one-loop & $E_{\textrm{f}}$ & -$E_{\textrm{f}}$ &  ?\\[2mm]
two-loop & $\mathcal{E}_{\scalebox{0.5}{EFT}}$ &  ? & $0$\\[2mm]
three-loop & $ E_{\textrm{f}}+ E_{\textrm{adj}}$ &0 &0
\end{tabular}
\end{center}
where the question marks stand for contributions we have not computed, and $\mathcal{E}_{\scalebox{0.5}{EFT}}$ is the two loop contribution from exceptional field theory computed in \cite{Bossard:2015foa}. For simplicity we have absorbed the numerical coefficients of these functions in their definition in the table, such that {\it e.g} $E_{\textrm{f}}=40\xi(2)\xi(6)\xi(d+3) E_{\alpha_d, \frac{d+3}{2}}$, where the label `f' stands for fundamental. Typically, the fundamental and adjoint functions are divergent at the values of their parameters (like $(d+3)/2$) and we employ minimal subtraction of the pole when continuing in this parameter to obtain regularised series $\hat{E}_{\textrm{f}}$ and $\hat{E}_{\textrm{adj}}$ and write the `hat' when we want to insist on this fact.

We therefore see from supersymmetry arguments that there is a potential contribution from 1/8 BPS states at one-loop that should correspond to a finite contribution compatible with the homogeneous differential equation satisfied by the threshold function~\cite{Green:2005ba,Bossard:2015uga}. 1/8 BPS black hole solutions in $D=4$ come in two forms: either with vanishing or with finite horizon area. As argued in~\cite{Bossard:2015foa}, relying on \cite{Papageorgakis:2014dma}, the finite size solitons should be exponentially suppressed in perturbation theory and therefore should not contribute to the low-energy effective action. This argument might be invalidated by the fact that they are expected to contribute with an exponentially growing multiplicity proportional to $B_{14}$. The vanishing size solitons should not arise separately from the generic 1/8 BPS solitons from the point of view of automorphic representations. Indeed, 4-dimensional solitons can be thought as 3-dimensional instantons \cite{Gunaydin:2005mx}, and there is no $E_8$ automorphic form with a wave front set of Bala--Carter type $3A_1$ that would not include   Fourier coefficient (intantons corrections) of Bala--Carter type $A_2$~\cite{Jiang:2014}, since the $3A_1$ orbit is not special. In simpler terms, there is no $E_8$ automorphic form that gets contributions from zero size 4D black holes but not from finite 4D black holes. In five space-time dimensions there are no vanishing size 1/8 BPS black hole  solitons, and there are no 1/8 BPS black hole solitons in $D\ge 6$. Since all contributions to the effective action come in a rather uniform way in all dimensions, it seems plausible that there is no one-loop contribution from 1/8 BPS solitons to the $\nabla^6 R^4$ coupling. One cannot justify this absence of contribution from the analysis of~\cite{Papageorgakis:2014dma} or by supersymmetry arguments, but we shall argue below that it must be the case if one assumes the loop expansion involving all BPS states to be consistent as an effective theory.

The two-loop exceptional field theory calculation in~\cite{Bossard:2015foa} gives a function  $\mathcal{E}_{\scalebox{0.5}{EFT}}$ that satisfies the correct tensorial inhomogeneous differential equation. The only automorphic solution to the homogeneous part of the tensorial differential equation for the non-chiral invariant is the adjoint Eisenstein series $E_{\textrm{adj}}$ discussed above. Since we have not analysed the perturbative limit of the function $\mathcal{E}_{\scalebox{0.5}{EFT}}$, we only know that it must reproduce the function $\mathcal{E}$ satisfying the same equation that appears in string theory,  modulo a free coefficient 
\be 
\mathcal{E}= \mathcal{E}_{\scalebox{0.5}{EFT}} - \alpha E_{\textrm{adj}} \ . 
\ee
The adjoint Eisenstein series does not have a perturbative string theory expansion along the T-duality group $O(d-1,d-1)$ that is compatible with string perturbation theory,\footnote{This means that they have powers of $g_\textrm{s}$ appearing that do not correspond to a positive integer genus calculation, most notably they contain na\"ive $-1/2$-loop order terms.} so one could in principle determine $\alpha$ by extracting the perturbative component in the string coupling constant of the exceptional field theory integral. It is important to note that these functions are divergent for $D=4,5,6$ and must be regularised appropriately. The logarithmic divergences of the string theory coupling $\mathcal{E}$ computed in \cite{Pioline:2015yea} coincide with the divergences of the adjoint Eisenstein series~\cite{Bossard:2015oxa}. They satisfy the same differential equations with the same linear inhomogeneous terms associated to the supergravity divergences. This implies in particular that their difference $\mathcal{E}- E_{\textrm{adj}}$ is well-defined and finite. 

The tree level supersymmetry Ward identities imply the homogeneous differential equations, and so the linear inhomogeneous terms must be compensated by the non-analytic component of the amplitude. In six dimensions the constant source term is associated to the 3-loop divergence, and so should only be present at 3-loop. In five dimensions the  linear source term is associated to the 2-loop form-factor of the exact $R^4$ coupling that appears at 1-loop. So once again it only contributes at 3-loop order. In four dimensions the linear source term is associated to the 1-loop form-factor of the exact $\nabla^4 R^4$ coupling. But we have seen in the preceding section that the $\nabla^4 R^4$ coupling comes entirely from the 2-loop contribution in our construction, so once again the source term to the differential equation only appears at 3-loop order. 

At three loops, we have calculated the 1/2 BPS contribution from exceptional field theory in this paper in~\eqref{eq:6DE3}, \eqref{eq:5DE3} and \eqref{4DE3}  and have found a combination of the adjoint function together with the other homogeneous solution ${E}_{\textrm{f}}$, so schematically 
\begin{align}
\mathcal{E}_{\gra{0}{1}}^{\scalebox{0.6}{3-loop}} =\hat{E}_{\textrm{adj}}+ \hat{E}_{\textrm{f}}\,
\end{align}
This combination was shown to be precisely such that its divergences compensate the ones of the non-analytic component of the 3-loop amplitude, and it satisfies the differential equations involving the inhomogeneous source terms consistent with the supergravity divergences discussed above~\cite{Bossard:2015oxa}.  Moreover, together with the contributions from lower loops one should obtain the correct full answer that has the schematic form
\begin{align}
\mathcal{E}_{\gra{0}{1}}^D = \hat{\mathcal{E}} + \hat{E}_{\textrm{f}}\,
\end{align}
as a combination of the regularised inhomogeneous solution and a regularised homogeneous solution. This is precisely consistent with the assumption that the amplitude should be finite to all orders, since the difference $\mathcal{E}- E_{\textrm{adj}}$ is finite and satisfies the inhomogeneous equation with a quadratic source term, but no linear source terms associated to threshold effects. We note that the fact that our homogeneous one-loop result $\hat{E}_{\textrm{f}}$ appears with exactly the same coefficient as in the proposal in~\cite{Pioline:2015yea} for $D=6$ suggests that $\alpha$ vanishes.

If we assume that the loop expansion involving all BPS states of the theory is indeed finite and well-defined order by order in the loop expansion, the couplings at 1-loop and 2-loop must be regular finite automorphic functions satisfying the differential equations with zero linear source terms associated to supergravity divergences. $\hat{E}_{\textrm{adj}}$ and $ \hat{E}_{\textrm{f}}$ satisfy differential equations with constant source terms, and their linear combination for which these source terms compensate for $D=4,5,6$ is not consistent with the decompactification limit. Indeed, using the formulas derived in~\cite{Bossard:2015oxa}, one can check that their appropriate linear combination such that the source terms compensate is $7 E_{\textrm{f}}- (10-d) E_{\textrm{adj}}$,\footnote{One computes that for $d=5$, $\Delta \hat{E}_{\textrm{adj}} = \frac{70}{3} \zeta(3) $ and  $\Delta \hat{E}_{\textrm{f}} = \frac{50}{3} \zeta(3) $; for $d=6$ that $(\Delta+18) \hat{E}_{\textrm{adj}} = \frac{35}{3} \mathcal{E}_{\gra{0}{0}} $ and  $(\Delta+18) \hat{E}_{\textrm{f}} = \frac{20}{3} \mathcal{E}_{\gra{0}{0}} $; and for $d=7$ that $(\Delta+60) \hat{E}_{\textrm{adj}} = \frac{35}{\pi} \mathcal{E}_{\gra{1}{0}} $ and  $(\Delta+60) \hat{E}_{\textrm{f}} = \frac{15}{2\pi} \mathcal{E}_{\gra{1}{0}} $.} whereas the normalisation must be independent of the dimension for them to be compatible in the decompactification limit, {\it i.e.} such that the large radius limit of the $D=4$ function reproduces at leading order the $D=5$ function, and idem from $D=5$ to $D=6$. So assuming moreover that the loop expansion involving all BPS states is compatible with the decompactification limit, one concludes that there cannot be any contribution from these functions to the $\nabla^6 R^4$ coupling at one loop, and that the 2-loop contribution must be precisely the well-defined finite combination  $\hat{\mathcal{E}}-\hat{E}_{\textrm{adj}} $. This implies that the 3-loop contribution to the $\nabla^6 R^4$ coupling we have computed in this paper is indeed the expected answer, despite the fact that it is by itself inconsistent with string perturbation theory. As the same inconsistency appears the the two-loop level with opposite sign, the overall perturbative answer is perfectly compatible with string theory. The result of this discussion can be summarised by the following table.
\begin{center}
\begin{tabular}{c|ccc}
$\nabla^6 R^4$ & $\frac12$ BPS & $\frac14$ BPS & $\frac18$ BPS\\[2mm]\hline\\[-2mm]
one-loop & $E_{\textrm{f}}$ & -$E_{\textrm{f}}$ & $0$ \\[2mm]
two-loop & $\mathcal{E} - \alpha E_{\textrm{adj}}$ &  $(\alpha-1) E_{\textrm{adj}}$ & $0$\\[2mm]
three-loop & $E_{\textrm{f}}+ E_{\textrm{adj}}$ &0 &0
\end{tabular}
\end{center}

Let us summarise the discussion of this section. By the supersymmetry arguments of~\cite{Bossard:2015uga} reviewed above there are only three types of functions that can arise in this table. Because of the quadratic source term, the function $\mathcal{E}$ can only appear once at 2-loop order. Considering in more detail the linear source terms associated to threshold effects one finds that the unique linear combination  of these three functions compatible with the decompactification limit that can appear before 3-loop order is  $\mathcal{E}-\hat{E}_{\textrm{adj}} $. We conclude that the total 1-loop contribution must vanish, the total 2-loop contribution must be $\mathcal{E}-\hat{E}_{\textrm{adj}} $, and the 3-loop contribution must be ${E}_{\textrm{f}}+ E_{\textrm{adj}}$ as we have obtained in this paper. This implies in turn that there should be no 1/8 BPS contributions at 1-loop order, and that the 2-loop contributions involving 1/4 BPS states should be proportional to the adjoint series with the correct coefficient. For this we have assumed that the loop expansion is finite and consistent with supersymmetry and the decompactification limit order by order in the loop expansion. It would be good to be able to confirm these predictions for the 1-loop contribution of 1/8 BPS states and the 2-loop contribution of 1/4 BPS states. For the second one could in principle apply the same tricks as in this paper, and U-dualise the 2-loop string theory amplitude contribution to the $\nabla^4 R^4$ coupling~\cite{DHoker:2005vch,DHoker:2014oxd}.

\subsection*{Acknowledgements}
We gratefully acknowledge discussions with M.~Patnaik, B.~Pioline and G.~Savin. This work was partially supported by a PHC PROCOPE, projet No 37667ZL, and by DAAD PPP grant 57316852 (XSUGRA). The work of G.B. was partially supported by the ANR grant Black-dS-String. We thank the Banff International Research Station for hospitality during part of this work.

\appendix

\section{Some matrix integrals and volumes of fundamental domains}
\label{app:matrixint}

For the determination of the three-loop diagram we require a number of identities of integrals over the space $H_{n\times n}^+$ of symmetric positive definite $(n\times n)$-matrices or quotients of this space. We collect these identities in this appendix.

The first identity is useful for the Schwinger $\Omega$-integrals and can be found for example in~\cite[(1.1)]{Herz}:
\begin{align}
\label{eq:MG}
\int\limits_{H_{n\times n}^+} \frac{d^{n(n+1)/2} \Omega}{\left(\det \Omega\right)^{\frac{n+1}2-s}}e^{-\pi \Tr(\Omega X)} =   \left|\det X\right|^{-s} \prod_{j=0}^{n-1} \pi^{-(s-\frac{j}2)}\Gamma\left(s-\frac{j}2\right)\,.
\end{align}
A similar identity for the $t$-type integrals in the text is of $\Gamma$-type (for $n>1$) and also involves the $\zeta$-function:
\begin{align}
\label{eq:hcos}
\int\limits_{H_{n\times n}^+/PGL(n,\mathds{Z})} \frac{d^{n(n+1)/2} t}{\left(\det t\right)^{\frac{n+1}2-s}} e^{-\pi \mu^2 \det t } &= \left( \prod_{j=2}^{n} \xi(j) \right) \cdot \int_0^\infty \frac{dt}{t^{1-s}} e^{-\pi t \mu^2} \nn\\
&=  (\pi\mu^2)^{-s} \Gamma(s) \prod_{j=2}^{n} \xi(j)  \,,
\end{align}
where we have separated out the determinant of $t$ and used the volume of unit determinant, symmetric, positive definite matrices up to action by $PGL(n,\mathds{Z})$ given by a product of completed Riemann zeta functions $\xi(k) = \pi^{-k/2} \Gamma(k/2)\zeta(k)$ given for example in~\cite[Sec. 4.4, Thm. 4]{Terras}.\footnote{We note that on the real line the function $\xi(k)$ has simple poles at $k=0$ and $k=1$ with residues $-1$ and $+1$, respectively. It also satisfies the functional relation $\xi(k)=\xi(1-k)$.} For our application we need the formula~\eqref{eq:MG} above for $X=M\tau_\gamma M^T$, where $M$ is summed over cosets of full rank matrices of the form
\begin{align}
&\quad \sum_{\substack{M\in \mathds{Z}^{n\times n}/GL(n,\mathds{Z})\\\det M\neq0}} \int\limits_{H_{n\times n}^+} \frac{d^{n(n+1)/2} \Omega}{\left(\det \Omega\right)^{\frac{n+1}2-s}}e^{-\pi \Tr(\Omega M\tau_\gamma M^T)} \nn\\
&=   (\det\tau_\gamma)^{-s}  \left( \prod_{j=0}^{n-1} \pi^{-(s-\frac{j}2)}\Gamma\left(s-\frac{j}2\right)\right)\sum_{\substack{M\in \mathds{Z}^{n\times n}/GL(n,\mathds{Z})\\ \det M \neq 0}} \left|\det M\right|^{-2s}\nn\\
&= (\det\tau_\gamma)^{-s}  \left( \prod_{j=0}^{n-1} \pi^{-(s-\frac{j}2)}\Gamma\left(s-\frac{j}2\right)\zeta(2s-j)\right)\nn\\
&= (\det\tau_\gamma)^{-s}   \prod_{j=0}^{n-1} \xi\left(2s-j\right)\,,
\end{align}
where we have used a special case of the Koecher zeta function in the next-to-last step~\cite[Sec. 4.4]{Terras}. An elementary way of understanding the appearance of the product of Riemann zeta functions is to use the property that a representative of each $GL(n,\mathds{Z})$ orbit is realized by an upper triangular matrix with generic positive diagonal entries $m_{ii}$ and off-diagonal entries $0\le m_{ij}<  m_{jj}$ for $i<j$. Since the determinant does not depend on the off-diagonal entries, they simply give an additional factor of $m_{ii}^{\, i-1}$, which after summing over $m_{ii}$ gives the  $\zeta$ function terms with increasing arguments.

\section{Ramanujan identity}
\label{app:RI}

We here provide a brief proof of the Ramanujan identity that appears in~\eqref{eq:RI}, using standard methods of Dirichlet series and Euler products. If a series $a(n)$ for $n\in \mathbb{N}$ is multiplicative, i.e., satisfies $a(mn) = a(m) a(n)$ whenever $m$ and $n$ are co-prime, one can express the corresponding Dirichlet series as an Euler product
\begin{align}
\sum_{n>0} a(n) n^{-s} = \prod_{p\,\textrm{prime}} P(p,s)\,,
\end{align}
where
\begin{align}
\label{eq:Bellseries}
P(p,s) = \sum_{k\geq 0} a(p^k) p^{-ks}\,.
\end{align}
The identity of the sum and product form follows formally from prime factorisation of integers.

The divisor sum $\sigma_a(n) = \sum_{d|n} d^a$ can easily be seen to be multiplicative. Moreover,
\begin{align}
\sigma_a(p^k) = \sum_{m=0}^k p^{ma}  = \frac{1-p^{a(k+1)}}{1-p^a}\,.
\end{align}
The product $\sigma_a(n)\sigma_b(n)$ is also multiplicative. The (Bell) series~\eqref{eq:Bellseries} becomes in this case
\begin{align}
\sum_{k \geq 0} \sigma_a(p^k) \sigma_b(p^k) p^{-ks} = \frac{1-p^{a+b-2s}}{(1-p^{-s})(1-p^{a-s})(1-p^{b-s})(1-p^{a+b-s})}\,.
\end{align}
Using the Euler product of the Riemann zeta series $\zeta(s) = \prod_{p} (1-p^{-s})^{-1}$ one therefore obtains
\begin{align}
\sum_{n>0} \sigma_a(n) \sigma_b(n) n^{-s} = \frac{\zeta(s)\zeta(s-a)\zeta(s-b)\zeta(s-a-b)}{\zeta(2s-a-b)}\,.
\end{align}
Putting $a=b=k$ one obtains~\eqref{eq:RI}.

\section{On affine Eisenstein series and Epstein series}
\label{app:E9}

Garland has studied in detail affine Eisenstein series and their functional relation~\cite{Garland1,Garland2}. We adapt his conventions and define an affine Eisenstein series for $E_9$ through an infinitesimal quasi-character given by a weight
\begin{align}
\label{eq:affwt}
\lambda = \sum_{i=1}^9 2s_i \Lambda_i -\rho + t \delta\,,
\end{align}
where $\rho=\sum_{i=1}^9 \Lambda_i$ is a standard choice of Weyl vector in terms of the fundamental weights $\Lambda_i$ and $\delta$ the primitive null root. For $E_9$ there are nine simple co-roots $h_i$ and one more Cartan subalgebra element that we denote by $d$~\cite{Kac}. We define the fundamental weights by
\begin{align}
\Lambda_i(h_j) = \delta_{ij} \quad\textrm{for $i,j,=1,\ldots,9$ and}\quad \Lambda_i(d)=0\,.
\end{align}
Simple Weyl reflections $w_i$ ($i=1,\ldots,9$) act on weights $\lambda$ by
\begin{align}
w_i(\lambda) = \lambda - \langle \lambda | \alpha_i \rangle \alpha_i\,,
\end{align}
where $\alpha_i$ are the simple roots. 

An Eisenstein series is now defined on elements $g$ of the centrally extended loop group and a variable $v$ associated with the direction $d$.\footnote{There are different definitions of groups in the Kac--Moody case. The minimal definition is as the group generated from the one-parameter subgroups associated with all the real roots~\cite{Peterson:1983}. The \textit{complete} Kac--Moody group is obtained by a certain completion with respect to a positive (or negative) Borel subgroup. In the affine case, the difference between these groups can be phrased as follows: The minimal definition corresponds to allowing only rational maps from $\mathds{C}^\times$ to the finite-dimensional Lie group (e.g., $E_8$) while the completed group allows infinite power series in the positive (or negative) powers of the variable on $\mathds{C}^\times$~\cite{Garland1,Garland2}. This complete group is the one that we are using here and should also be the one that is relevant in two-dimensional supergravity~\cite{Breitenlohner:1986um}. } In other words, torus elements are written as $a v^d$ for $a=\prod_{i=1}^9 r_i^{h_i}$ and the pairing with the weight $\lambda$ of~\eqref{eq:affwt} is by
\begin{align}
\exp \langle \lambda+\rho  | H(av^d)\rangle = (a v^d)^{\lambda+\rho} = v^t \prod_{i=1}^9 r_i^{2s_i}\,,
\end{align}
where $H$ denotes the logarithm map from the affine group to the split torus. The Eisenstein series is then given by
\begin{align}
E(\lambda,gv^d) = \sum_{\gamma\in \hat{B}(\mathds{Z})\backslash G(\mathds{Z})} e^{\langle \lambda+\rho | H(\gamma g v^d) \rangle} = \sum_{\gamma\in \hat{B}(\mathds{Z})\backslash G(\mathds{Z}) }\gamma\left[  (gv^d)^{\lambda+\rho}\right]
\end{align}
as a sum over a discrete group of the centrally extended loop group, \ie the affine group $E_9$ without the $d$-direction. Convergence requires restricting also the group element $gv^d$, in particular the $v$ coordinate~\cite{Garland1}.

The functional relation for Weyl related weights $\lambda$ and $w\lambda$ is 
\begin{align}
\label{eq:FR}
E(\lambda,gv^d) = M(w,\lambda) E(w\lambda,gv^d)
\end{align}
for 
\begin{align}
M(w,\lambda) = \prod_{\alpha>0\atop w\alpha<0} \frac{\xi(\langle \lambda | \alpha \rangle)}{\xi(\langle \lambda | \alpha \rangle+1)}
\end{align}
as usual. An important point now is that $w\lambda$ can also alter the coefficient of $\delta$ in~\eqref{eq:affwt} thus changing the overall power of $v$. This explains the different powers of $v$ appearing in relations such as~\eqref{eq:affFR1}.

From the point of view of a putative $E_9$ exceptional field theory, some of the $E_9$ Eisenstein series should arise from Feynman diagrams. The coordinates are expected to lie in the highest weight representation representation $R(\Lambda_9+\delta)$~\cite{Bossard:2017aae}, such that the dual discrete charges $\Gamma$ are in the conjugate $\overline{R(\Lambda_9+\delta)}$. The shift in $\delta$ corresponds to factors of $v$ appearing in the BPS mass and also follows from the decomposition of the fundamental representation $R(\Lambda_{11})$ of $E_{11}$~\cite{West:2003fc,Kleinschmidt:2003jf}.\footnote{One useful observation for checking this is that $\delta$ is related to minus $\Lambda_{10}$.} Considering the non-linear sigma model with the three-dimensional metric 
\be 
ds^2 = e^{2\sigma}( - dt^2 + dx^2 ) + v^2 dy^2 \ , 
\ee
with $y$ a circle coordinate, the $E_{8(8)}$ non-linear sigma model action involves the Lagrangian 
\be 
\cL = v \Tr ( P_t P_t - P_x P_x - ( v^{-1} P_y ) ( v^{-1} P_y )  ) \ . 
\ee
The second term implies indeed that, after reduction to two space-time dimensions, the mass formula for a charge $\Gamma \in \overline{R(\Lambda_9+\delta)}$ is $v^{-1} | Z(\Gamma)|$ where $Z(\Gamma)$ is defined in the representation $\overline{R(\Lambda_9)}$. In the Feynman amplitude, one must also take into account the power of $v$ due to the overall $v$ factor in the Lagrangian. Each vertex contributes a factor of $v$, and each internal line a factor of $v^{-1}$, such that at $L$-loop one gets an overall factor of $v^{1-L}$. 

The relevant lattice in $\overline{R(\Lambda_9+\delta)}$ in which the charge $\Gamma$ is defined is the smallest lattice in $\overline{R(\Lambda_9+\delta)}$ that includes the canonical lowest weight representative vector $\Gamma_0 $ with an arbitrary integer coefficient and that is preserved by the Chevalley group  $E_9(\mathds{Z})$. Its elements are linear combinations over $\mathds{Z}$ of charge vectors obtained from  the canonical lowest weight representative by the action of the Chevalley group $E_9(\mathds{Z})$. The same definition applies for the module defined over $\mathds{Q}$. According to \cite{Peterson:1983}, the solution to the constraint  $ \Gamma\times \Gamma=0$ in the module $\overline{R(\Lambda_9+\delta)}$ over $\mathds{Q}$ defines a single orbit of $E_9(\mathds{Q})$ of the canonical lowest weight vector representative. Because of the Bruhat decomposition of $E_9(\mathds{Q})$
\be E_9(\mathds{Q}) =  \bigcup_{w\in W} B(\mathds{Q}) w B(\mathds{Q})    \ , \ee
one can prove recursively using the property that $SL(2,\mathds{Q}) = SL(2,\mathds{Z}) B(\mathds{Q})$ that 
\begin{align} 
\label{eq:E9dec}
 E_9(\mathds{Q}) = E_9(\mathds{Z}) B(\mathds{Q}) \  , 
 \end{align}
where $B(\mathds{Q}) $ is the Borel subgroup of $E_9(\mathds{Q})$. Let us sketch a proof of this statement that is standard for finite-dimensional simple groups. Bruhat decomposition implies that any element $g\in G(\mathds{Q})$ belonging to a given Bruhat cell $B(\mathds{Q})w B(\mathds{Q})$ is the product of certain elements $g_\alpha$ for each positive root $\alpha \in \Delta^+$ in certain $SL(2,\mathds{Q})_\alpha$ subgroups associated to a Weyl word $w$. More explicitly, let $w=w_{i_\ell}\cdots w_{i_1}$ be a reduced expression of a Weyl word of length $\ell$ in terms of simple reflections such that $w^{-1} = w_{i_1} \cdots w_{i_\ell}$. Then, for $k=1,\ldots,\ell$,  the negative roots
\begin{align}
\beta_k= w_{i_1} \cdots w_{i_{k}} \alpha_{i_{k}}
\end{align}
with $\alpha_{i_k}$ the $i_k$th simple root parametrise all roots $\alpha<0$ such that $w \alpha>0$ (see, e.g., \cite{Kumar}). In particular $\beta_1=-\alpha_{i_1}$. Any element $g\in B(\mathds{Q})w B(\mathds{Q})$ can then be written uniquely as 
\be 
g  = w \left[\prod_{k=\ell}^1 \bar{n}_{\beta_k}\right] b \
\ee
with $\bar{n}_{\beta_k} \in \bar{N}_{\beta_k}=\left\{ \exp(q E_{ \beta_k})\,|\,q\in\mathds{Q}\right\} $, the lower unipotent inside $SL(2,\mathds{Q})_{\beta_k}$ and $b\in B(\mathds{Q})$. We choose to order the factors starting with $\beta_\ell$ on the left as indicated by the limits on the product. Note that we can think of the Weyl word $w$ as an element of $E_9(\mathds{Z})$. The equality $SL(2,\mathds{Q}) = SL(2,\mathds{Z}) B(\mathds{Q})$ then implies that we can write\footnote{Explicitly, we have for co-prime $p$ and $q$
\begin{align}
\begin{pmatrix}1& 0 \\p/q&1\end{pmatrix} = \begin{pmatrix}q & b\\ p&d\end{pmatrix}\begin{pmatrix} 1/q & -b\\ 0 &q\end{pmatrix}\ , \nn 
\end{align}
where $b,d\in \mathds{Z}$ are any solution to $qd-pb=1$. The ambiguity in this solution can be absorbed in the Borel matrix on the right. We do not require this explicit form, however, for the argument.}
\be 
\bar{n}_{\beta_\ell} = h_{\beta_\ell} b_{\beta_\ell} \ , 
\ee
with $h_{\beta_\ell} \in SL(2,\mathds{Z})_{\beta_\ell}$ and  $b_{\beta_\ell} \in B_{\beta_\ell}(\mathds{Q})$, and then
\be 
g  =  w h_{\beta_\ell} \left[b_{\beta_\ell} \prod_{i=\ell-1}^1 \bar{n}_{\beta_i}\right] b  = wh_{\beta_\ell} \left[\prod_{i=\ell-1}^1   ( b_{\beta_\ell}  \bar{n}_{\beta_i} b_{\beta_\ell}^{-1} )\right]   b_{\gamma_\ell} b   =  wh_{\gamma_1} \left[\prod_{i=\ell-1}^1   \bar{n}_{\beta_i}^\prime \right] b'
\ee
with $b^\prime \in B(\mathds{Q})$ and new elements $\bar{n}_{\beta_i}^\prime$ in the same roots spaces. The last statement is true because the Borel conjugation with $b_{\gamma_\ell}$ scales an element $\bar{n}_{\beta_i}$ and produces contributions to $\bar{n}_{\beta_j}$ for $j<i$ (according to our ordering assumptions) as well as elements in the Borel $B(\mathds{Q})$. In $b'$ we have collected all possible such contributions. Recursively one gets  
\be g  =w\left[ \prod_{i=\ell}^1 h_{\beta_i} \right]\tilde{b}
\ee
with $\tilde{b}\in B(\mathds{Q})$. Thus any element $g\in BwB$ can be written as $E_9(\mathds{Z}) B(\mathds{Q})$. Since $E_9(\mathds{Q})$ is the union of its Borel cells and the statement is true for all Weyl words $w$, we obtain the claim~\eqref{eq:E9dec}. A similar argument that is also valid for arbitrary (completed) Kac--Moody group can be constructed using Theorem 8.15 of~\cite{Steinberg} that directly gives representatives for the Bruhat cells of the desired form.\footnote{We are grateful to M.~Patnaik and G.~Savin for explaining this general proof to us.}

Since the lowest weight vector is by definition stabilised by the conjugate (lower) Borel subgroup $B(\mathds{Q})^T$ up to the multiplication by a rational number, it follows that any solution to the constraint   $ \Gamma\times \Gamma=0$ in the discrete lattice in $\overline{R(\Lambda_9+\delta)}$ is in the $E_9(\mathds{Z})$ orbit of a canonical lowest weight representative  $\Gamma_0 = \mbox{gcd}(\Gamma) |\Lambda_9\rangle$.  
Such elements of the above lattice are associated to 1/2-BPS multiplets of states, and the 1-loop amplitude for four massless scalar fields in the putative two dimensional exceptional field theory are therefore formally Epstein series of the form
\begin{align}
\label{eq:oneloop9}
E(2s\Lambda_9+2s\delta-\rho) =\frac{1}{2\zeta(2s)} v^{2s} \sump_{\Gamma \in \mathds{Z}^{d(\alpha_9)}\atop \Gamma\times \Gamma=0} |Z(\Gamma)|^{-2s}\,,
\end{align}
where the coefficient of $\delta$ is fixed by the shifted representation of the charges. On the right-hand side, the charges are given in the standard lattice in $\overline{R(\Lambda_9)}$ and we have written the $\delta$ shift by extracting the power of $v^{2s}$. This formal definition defines an absolutely convergent sum provided the moduli are restricted to (a slice of) the Tits cone in $K(E_9)\backslash E_{9(9)}$ and the $\mbox{Re}(s)$ is sufficiently large, and is then mathematically sound.  We expect a similar rewriting of the Langlands Eisenstein series as an Epstein for general Kac--Moody algebras, but note that the convergence and analytic continuation have not been established rigorously, see~\cite{Carbone:2017} for recent results. The functional relation of the Langlands Eisenstein series was proven by Garland in~\cite{Garland1,Garland2}.

These powers of $v$ and coefficients are consistent with the results found in~\cite{Fleig:2012xa} when applying an appropriate functional relation~\eqref{eq:FR}. More precisely, we have from~\cite[Eq.~(3.9)]{Bossard:2015foa} at one loop the following threshold functions
\begin{align}
R^4:&& 4\pi \xi(6) E(2\cdot 3\Lambda_9+6\delta-\rho) &= 4\pi \xi(3) E(2 \cdot \tfrac32 \Lambda_1 +\delta-\rho) = 2\zeta(3) v E(2 \cdot \tfrac32 \Lambda_1 -\rho) \,,\nn\\
\nabla^4 R^4:&& \frac{4\pi^3}{45} \xi(10) E(2\cdot 5\Lambda_9+10\delta-\rho) &=\frac{4\pi^3\xi(2)\xi(5)}{45\xi(4)} E(2 \cdot \tfrac52 \Lambda_1 +\delta-\rho) = \zeta(5) v E(2 \cdot \tfrac52 \Lambda_1 -\rho) 
\end{align}
and these are exactly the functions with the correct $v$ powers found in~\cite{Fleig:2012xa}.

Let us also consider formally the two- and three-loop amplitudes to be constructed $E_9$ exceptional field theory. The 2-loop contribution to the $\nabla^4 R^4$ coupling computed in~\cite[Eq.~(4.47)]{Bossard:2015foa}  defines an Epstein sum over the wedge product of the two 1/2-BPS charges, that give rise to an Epstein sum in the largest module $R(\Lambda_{d-1})$ in the wedge product $R(\Lambda_{d})\wedge R(\Lambda_{d}) $. Similarly, the 3-loop contribution to $\nabla^6 R^4$ computed in \eqref{eq:a4} gives an Epstein sum in the largest module $R(\Lambda_{d-2})$ in the 3-form wedge product $R(\Lambda_{d})\wedge R(\Lambda_{d}) \wedge R(\Lambda_d)$. Applying the same argument for $E_9$, and taking into account the $\delta$ shift, one obtains  
\begin{align}
R(\Lambda_9+t \delta ) \wedge R(\Lambda_9+t \delta) &= R(\Lambda_8+(2t-1) \delta ) \oplus\ldots\,,\nn\\
R(\Lambda_9+t \delta) \wedge R(\Lambda_9+t \delta) \wedge R(\Lambda_9+t \delta) &= R(\Lambda_7+(3t-2) \delta) \oplus\ldots\, ,
\end{align}
such that $L(t-1)+1=1$ for $R(\Lambda_9+ \delta )$ and one expects the $L$-loop contribution in exceptional field theory to give a contribution proportional to the Eisenstein series  $v^{1-L} E(2s(\Lambda_{10-L} + \delta)-\rho)= E(2s \Lambda_{10-L} + (2s-L+1)\delta -\rho)$ for the appropriate value of $s$. In other words, the correct function at two loops should be with weight $\lambda = 2\cdot \frac52\Lambda_8 +(2\cdot\tfrac52-1) \delta-\rho$ and for three-loops with weight $\lambda = 2\cdot 2\Lambda_7 +(2\cdot 2-2) \delta-\rho$, where the former corresponds to the $\nabla^4R^4$ threshold arising at two loops~\cite[Eq.~(4.47)]{Bossard:2015foa} and the latter to the homogeneous solution for the $\nabla^6 R^4$ threshold arising at three-loops~\eqref{eq:a4}. 

We can verify these expectations by applying functional relations of the type~\eqref{eq:FR}. For the $\nabla^4 R^4$ threshold function one finds 
\begin{align}
8\pi \xi(4)\xi(5) E(5 \Lambda_8 +4\delta -\rho) = 8\pi \xi(2)\xi(5)E(2\cdot \tfrac52 \Lambda_1 +\delta -\rho) = \zeta(5) v E(2\cdot\tfrac52\Lambda_1-\rho)
\end{align}
as required for $\nabla^4R^4$~\cite{Fleig:2012xa}.

The homogeneous solution for $\nabla^6 R^4$ deduced in this paper from the three-loop calculation is
\begin{align}
40\xi(2)\xi(3)\xi(4)E(2\cdot 2 \Lambda_7 +2\delta -\rho) = 40\xi(2)\xi(6)\xi(12)E(2\cdot 6 \Lambda_8+12 \delta-\rho)\,,
\end{align}
which is the result found at one-loop in \cite{Bossard:2015foa} with~\eqref{eq:oneloop9} at $s=6$.


\begin{thebibliography}{99}


\bibitem{Green:1981yb}
  M.~B.~Green and J.~H.~Schwarz,
  ``Supersymmetrical String Theories,''
  \doi{Phys.\ Lett.\  {\bf 109B} (1982) 444}{doi:10.1016/0370-2693(82)91110-8}.

\bibitem{GrossWitten}
  D.~J.~Gross and E.~Witten, 
  ``Superstring Modifications of Einstein's Equations,'' 
  \doi{Nucl.Phys. B277 (1986) 1}{http://dx.doi.org/10.1016/0550-3213(86)90429-3}.

\bibitem{Green:1997tv}
  M.~B.~Green and M.~Gutperle,
  ``Effects of D instantons,''
  \doi{Nucl.\ Phys.\ B {\bf 498} (1997) 195}{doi:10.1016/S0550-3213(97)00269-1}
  \eprint{hep-th/9701093}.



  \bibitem{Kiritsis:1997em} 
  E.~Kiritsis and B.~Pioline,
  ``On $R^4$ threshold corrections in IIB string theory and $(p, q)$ string instantons,''
  Nucl.\ Phys.\ B {\bf 508} (1997) 509, 
  \eprint{hep-th/9707018}.


  
\bibitem{Obers:1999um}
  N.~A.~Obers and B.~Pioline,
  ``Eisenstein series and string thresholds,''
  Commun.\ Math.\ Phys.\  {\bf 209} (2000) 275, 
  \eprint{hep-th/9903113}.

\bibitem{Green:1999pv}
  M.~B.~Green and P.~Vanhove,
  ``The Low-energy expansion of the one loop type II superstring amplitude,''
  \doi{Phys.\ Rev.\ D {\bf 61} (2000) 104011}{doi:10.1103/PhysRevD.61.104011}
  \eprint{hep-th/9910056}.
 

\bibitem{deWit:1999ir}
  B.~de Wit and D.~L\"{u}st,
  ``BPS amplitudes, helicity supertraces and membranes in M theory,''
  Phys.\ Lett.\ B {\bf 477} (2000) 299, 
  \eprint{hep-th/9912225}.


\bibitem{Green:2005ba}
  M.~B.~Green and P.~Vanhove,
  ``Duality and higher derivative terms in M theory,''
  JHEP {\bf 0601} (2006)  093, 
  \eprint{hep-th/0510027}.
 
\bibitem{Basu:2007ck} 
  A.~Basu,
  ``The $D^6 R^4$ term in type IIB string theory on $T^2$ and U-duality,''
  Phys.\ Rev.\ D {\bf 77}  (2008) 106004, 
 \eprintN{0712.1252}.

\bibitem{Basu:2007ru}
  A.~Basu,
  ``The $D^4 R^4$ term in type IIB string theory on $T^2$ and U-duality,''
  Phys.\ Rev.\ D {\bf 77} (2008)   \href{http://dx.doi.org/10.1103/PhysRevD.77.106003}{106003}, \eprintN{0708.2950}.
 

\bibitem{Green:2008uj}
  M.~B.~Green, J.~G.~Russo and P.~Vanhove,
  ``Low energy expansion of the four-particle genus-one amplitude in type II superstring theory,''
  \doi{JHEP {\bf 0802} (2008) 020}{doi:10.1088/1126-6708/2008/02/020}
  \eprintN{0801.0322}
 
\bibitem{Green:2010wi}
  M.~B.~Green, J.~G.~Russo and P.~Vanhove,
  ``Automorphic properties of low energy string amplitudes in various dimensions,''
  Phys.\ Rev.\ D {\bf 81} (2010) 086008, 
  \eprintN{1001.2535}.

\bibitem{Pioline:2010kb}
  B.~Pioline,
  ``$R^4$ couplings and automorphic unipotent representations,''
  JHEP {\bf 1003} (2010) 116, 
  \eprintN{1001.3647}.
    
\bibitem{Green:2010kv}
  M.~B.~Green, S.~D.~Miller, J.~G.~Russo and P.~Vanhove,
  ``Eisenstein series for higher-rank groups and string theory amplitudes,''
  Commun.\ Num.\ Theor.\ Phys.\  {\bf 4}  (2010)  551,  
\eprintN{1004.0163}.
 

\bibitem{Green:2011vz}
  M.~B.~Green, S.~D.~Miller and P.~Vanhove,
  ``Small representations, string instantons, and Fourier modes of Eisenstein series (with an appendix by D. Ciubotaru and P. Trapa),'' 
  J. Number Theory {\bf 146} (2015) 187--309, 
  \eprintN{1111.2983}.

\bibitem{Fleig:2012xa}
  P.~Fleig and A.~Kleinschmidt,
  ``Eisenstein series for infinite-dimensional U-duality groups,''
  \doi{JHEP {\bf 1206} (2012) 054}{doi:10.1007/JHEP06(2012)054}
  \eprintN{1204.3043}.

\bibitem{Bossard:2014lra}
  G.~Bossard and V.~Verschinin,
  ``Minimal unitary representations from supersymmetry,''
  \doi{JHEP {\bf 1410} (2014) 008}{doi:10.1007/JHEP10(2014)008}
  \eprintN{1406.5527}.

  
  \bibitem{Bossard:2014aea}
  G.~Bossard and V.~Verschinin,
  ``$\mathcal{E} \nabla^4 R^4$ type invariants and their gradient expansion,''
  \doi{JHEP {\bf 1503} (2015) 089}{doi:10.1007/JHEP03(2015)089}
  \eprintN{1411.3373}.


\bibitem{Pioline:2015yea}
  B.~Pioline,
  ``$D^{6}R^{4}$ amplitudes in various dimensions,''
  JHEP {\bf 1504} (2015) 057, 
  \eprintN{1502.03377}.

 \bibitem{Bossard:2015uga}
  G.~Bossard and V.~Verschinin,
  ``The two $\nabla^{6} R^{4}$ type invariants and their higher order generalisation,''
  \doi{JHEP {\bf 1507} (2015) 154}{doi:10.1007/JHEP07(2015)154}
  \eprintN{1503.04230}.
  
 \bibitem{Basu:2015dqa}
  A.~Basu,
  ``Perturbative type II amplitudes for BPS interactions,''
  \eprintN{1510.01667}

\bibitem{Hull:1994ys}
  C.~M.~Hull and P.~K.~Townsend,
  ``Unity of superstring dualities,''
  \doi{Nucl.\ Phys.\ B {\bf 438} (1995) 109}{doi:10.1016/0550-3213(94)00559-W}
  \eprint{hep-th/9410167}.



\bibitem{Green:1998by}
  M.~B.~Green and S.~Sethi,
  ``Supersymmetry constraints on type IIB supergravity,''
  \doi{Phys.\ Rev.\ D {\bf 59} (1999) 046006}{doi:10.1103/PhysRevD.59.046006}
  \eprint{hep-th/9808061}.

\bibitem{Pioline:1998mn}
  B.~Pioline,
  ``A Note on nonperturbative $R^4$ couplings,''
  \doi{Phys.\ Lett.\ B {\bf 431} (1998) 73}{doi:10.1016/S0370-2693(98)00554-1}
  \eprint{hep-th/9804023}.

\bibitem{Mizoguchi:1999fu}
  S.~Mizoguchi and G.~Schroder,
  ``On discrete U duality in M theory,''
  \doi{Class.\ Quant.\ Grav.\  {\bf 17} (2000) 835}{doi:10.1088/0264-9381/17/4/308}
  \eprint{hep-th/9909150}.

\bibitem{Green:2014yxa}
  M.~B.~Green, S.~D.~Miller and P.~Vanhove,
  ``$SL(2, \mathbb{Z})$-invariance and D-instanton contributions to the $D^6 R^4$ interaction,''
  \doi{Commun.\ Num.\ Theor.\ Phys.\  {\bf 09} (2015) 307}{doi:10.4310/CNTP.2015.v9.n2.a3}
  \eprintN{1404.2192}.

\bibitem{Basu:2014hsa}
  A.~Basu,
  ``The $D^{6}R^{4}$ term from three loop maximal supergravity,''
  \doi{Class.\ Quant.\ Grav.\  {\bf 31} (2014) no.24,  245002}{doi:10.1088/0264-9381/31/24/245002}
  \eprintN{1407.0535}.




\bibitem{Bossard:2015oxa}
  G.~Bossard and A.~Kleinschmidt,
  ``Supergravity divergences, supersymmetry and automorphic forms,''
  \doi{JHEP {\bf 1508} (2015) 102}{doi:10.1007/JHEP08(2015)102}
  \eprintN{1506.00657}.


\bibitem{Bossard:2015foa}
  G.~Bossard and A.~Kleinschmidt,
  ``Loops in exceptional field theory,''
  \doi{JHEP {\bf 1601} (2016) 164}{doi:10.1007/JHEP01(2016)164},
  \eprintN{1510.07859}.

\bibitem{Koepsell:2000xg}
  K.~Koepsell, H.~Nicolai and H.~Samtleben,
  ``An exceptional geometry for D = 11 supergravity?,''
  Class.\ Quant.\ Grav.\  {\bf 17} (2000) 3689, 
  \eprint{hep-th/0006034}.
 
\bibitem{Hull:2007zu}
  C.~M.~Hull,
  ``Generalised geometry for M-theory,''
  JHEP {\bf 0707} (2007) 079, 
  \eprint{hep-th/0701203}.

\bibitem{Pacheco:2008ps}
  P.~P.~Pacheco and D.~Waldram,
  ``M-theory, exceptional generalised geometry and superpotentials,''
  JHEP {\bf 0809} (2008) 123, 
  \eprintN{0804.1362}.
 
\bibitem{Berman:2010is}
  D.~S.~Berman and M.~J.~Perry,
  ``Generalized geometry and M theory,''
  JHEP {\bf 1106} (2011) 074, 
  \eprintN{1008.1763}.
  
\bibitem{Berman:2011cg}
  D.~S.~Berman, H.~Godazgar, M.~Godazgar and M.~J.~Perry,
  ``The local symmetries of M-theory and their formulation in generalised geometry,''
  JHEP {\bf 1201} (2012) 012, 
 \eprintN{1110.3930}.

\bibitem{Cederwall:2013naa}
  M.~Cederwall, J.~Edlund and A.~Karlsson,
  ``Exceptional geometry and tensor fields,''
  JHEP {\bf 1307} (2013) 028, 
  \eprintN{1302.6736}.
  
\bibitem{Hohm:2013pua} 
  O.~Hohm and H.~Samtleben,
  ``Exceptional form of $D=11$ Supergravity,''
  Phys.\ Rev.\ Lett.\  {\bf 111} (2013) 231601,  
  \eprintN{1308.1673}.
  
  
  
\bibitem{Hohm:2013vpa} 
  O.~Hohm and H.~Samtleben,
  ``Exceptional field theory I: $E_{6(6)}$ covariant form of M-theory and type IIB,''
  Phys.\ Rev.\ D {\bf 89} (2014) 066016, 
  \eprintN{1312.0614}.
  
  
\bibitem{Hohm:2013uia} 
  O.~Hohm and H.~Samtleben,
  ``Exceptional field theory. II. $E_{7(7)}$,''
  Phys.\ Rev.\ D {\bf 89} (2014) 066017, 
  \eprintN{1312.4542}.
  
 \bibitem{Aldazabal:2013via}
  G.~Aldazabal, M.~Gra\~na, D.~Marqu\'es and J.~A.~Rosabal,
  ``The gauge structure of exceptional field theories and the tensor hierarchy,''
  JHEP {\bf 1404} (2014) 049, 
  \eprintN{1312.4549}.
 
  
\bibitem{Godazgar:2014nqa}
  H.~Godazgar, M.~Godazgar, O.~Hohm, H.~Nicolai and H.~Samtleben,
  ``Supersymmetric $E_{7(7)}$ exceptional field theory,''
  JHEP {\bf 1409} (2014) 044, 
  \eprintN{1406.3235}.
  
\bibitem{Hohm:2014fxa}
  O.~Hohm and H.~Samtleben,
  ``Exceptional field theory. III. $E_{8(8)}$,''
  Phys.\ Rev.\ D {\bf 90} (2014) 066002, 
 \eprintN{1406.3348}.



\bibitem{Bern:2007hh}
  Z.~Bern, J.~J.~Carrasco, L.~J.~Dixon, H.~Johansson, D.~A.~Kosower and R.~Roiban,
  ``Three-loop superfiniteness of N=8 supergravity,''
  \doi{Phys.\ Rev.\ Lett.\  {\bf 98} (2007) 161303}{doi:10.1103/PhysRevLett.98.161303},
  \eprint{hep-th/0702112}.
  
\bibitem{Bern:2008pv}
  Z.~Bern, J.~J.~M.~Carrasco, L.~J.~Dixon, H.~Johansson and R.~Roiban,
  ``Manifest ultraviolet behavior for the three-loop four-point amplitude of N=8 supergravity,''
  \doi{Phys.\ Rev.\ D {\bf 78} (2008) 105019}{doi:10.1103/PhysRevD.78.105019},
  \eprintN{0808.4112 [hep-th]}.

\bibitem{Green:1997as}
  M.~B.~Green, M.~Gutperle and P.~Vanhove,
  ``One loop in eleven-dimensions,''
  \doi{Phys.\ Lett.\ B {\bf 409} (1997) 177}{doi:10.1016/S0370-2693(97)00931-3}
  \eprint{hep-th/9706175}.

\bibitem{Green:1999pu}
  M.~B.~Green, H.~h.~Kwon and P.~Vanhove,
  ``Two loops in eleven-dimensions,''
  \doi{Phys.\ Rev.\ D {\bf 61} (2000) 104010}{doi:10.1103/PhysRevD.61.104010}
  \eprint{hep-th/9910055}.


\bibitem{Gomez:2013sla} 
  H.~Gomez and C.~R.~Mafra,
  ``The closed-string 3-loop amplitude and S-duality,''
  \doi{JHEP {\bf 1310}, 217 (2013)}{doi:10.1007/JHEP10(2013)217}
  \eprintN{1308.6567}.
  
 \bibitem{Krutelevich}
  S.~Krutelevich, 
  ``Jordan algebras, exceptional groups, and Bhargava composition,''
   J.~Algebra {\bf 314} no. 2 (2007) 92. 
   

\bibitem{Bossard:2016hgy}
  G.~Bossard and B.~Pioline,
  ``Exact $\nabla^4 R^4$ couplings and helicity supertraces,''
  \doi{JHEP {\bf 1701} (2017) 050}{doi:10.1007/JHEP01(2017)050}
 \eprintN{1610.06693 [hep-th]}.




\bibitem{Julia}
  B. Julia, 
  {\it Kac-Moody symmetry of gravitation and supergravity theory}, 
  in Lectures in Applied Mathematics, AMS-SIAM, vol. 21 (1985), p. 35.

\bibitem{West:2001as}
  P.~C.~West,
  ``E$_{11}$ and M theory,''
  \doi{Class.\ Quant.\ Grav.\  {\bf 18} (2001) 4443}{doi:10.1088/0264-9381/18/21/305}
  \eprint{hep-th/0104081}.

\bibitem{Damour:2002cu}
  T.~Damour, M.~Henneaux and H.~Nicolai,
  ``E$_{10}$ and a 'small tension expansion' of M theory,''
  \doi{Phys.\ Rev.\ Lett.\  {\bf 89} (2002) 221601}{doi:10.1103/PhysRevLett.89.221601}
  \eprint{hep-th/0207267}.

\bibitem{Nicolai:1987kz}
  H.~Nicolai,
  ``The Integrability of $N=16$ Supergravity,''
  \doi{Phys.\ Lett.\ B {\bf 194} (1987) 402}{doi:10.1016/0370-2693(87)91072-0}.

\bibitem{Bossard:2017aae}
  G.~Bossard, M.~Cederwall, A.~Kleinschmidt, J.~Palmkvist and H.~Samtleben,
  ``Generalised diffeomorphisms for E$_9$,''
  \doi{Phys.\ Rev.\ D {\bf 96} (2017) 106022}{doi:10.1103/PhysRevD.96.106022}
  \eprintN{1708.08936}.
   
\bibitem{Garland1}
   H.~Garland,
   ``Certain Eisenstein series on loop groups: convergence and the constant term,''
   in: Algebraic groups and arithmetic, 275--319, Tata Inst. Fund. Res., Mumbai, 2004.

\bibitem{Garland2}
  H.~Garland, 
  ``Eisenstein series on arithmetic quotients of loop groups,''
   Math. Res. Lett. {\bf 6} (1999), 723--733. 
 
\bibitem{Carbone:2017}
  L.~Carbone, K.-H.~Lee and D.~Liu,
 ``Eisenstein series on rank 2 hyperbolic Kac-Moody groups,''
\doi{Math. Ann. {\bf 367} (2017) 1173--1197}{doi:10.1007/s00208-016-1428-8}.

\bibitem{Angelantonj:2012gw}
  C.~Angelantonj, I.~Florakis and B.~Pioline,
  ``One-Loop BPS amplitudes as BPS-state sums,''
   \doi{JHEP {\bf 1206} (2012) 070}{doi:10.1007/JHEP06(2012)070}, 
  \eprintN{1203.0566}.

\bibitem{Zagier:1981}
  D. Zagier, 
  ``The Rankin-Selberg method for automorphic functions which are not of rapid decay,''
  J. Fac. Sci. Univ. Tokyo Sect. IA Math. {\bf 28} (1981) 415--437 (1982).

\bibitem{Angelantonj:2011br}
  C.~Angelantonj, I.~Florakis and B.~Pioline,
  ``A new look at one-loop integrals in string theory,''
  \doi{Commun.\ Num.\ Theor.\ Phys.\  {\bf 6} (2012) 159}{doi:10.4310/CNTP.2012.v6.n1.a4}
  \eprintN{1110.5318}.
   
\bibitem{Obers:1998fb}
  N.~A.~Obers and B.~Pioline,
  ``U duality and M theory,''
  Phys.\ Rept.\  {\bf 318} (1999) 113, 
  \eprint{hep-th/9809039}.

\bibitem{Kiritsis:1997hj} 
  E.~Kiritsis,
  ``Introduction to superstring theory,''
  \eprint{hep-th/9709062}.


\bibitem{Gunaydin:2005mx} 
  M.~Gunaydin, A.~Neitzke, B.~Pioline and A.~Waldron,
  ``BPS black holes, quantum attractor flows and automorphic forms,''
  \doi{Phys.\ Rev.\ D {\bf 73}, 084019 (2006)}{doi:10.1103/PhysRevD.73.084019}
  \eprint{hep-th/0512296}.
  
\bibitem{Bossard:2016zdx}
  G.~Bossard, C.~Cosnier-Horeau and B.~Pioline,
  ``Protected couplings and BPS dyons in half-maximal supersymmetric string vacua,''
  \doi{Phys.\ Lett.\ B {\bf 765} (2017) 377}{doi:10.1016/j.physletb.2016.12.035}
  \eprintN{1608.01660}.
  
\bibitem{Bossard:2017wum}
  G.~Bossard, C.~Cosnier-Horeau and B.~Pioline,
  ``Four-derivative couplings and BPS dyons in heterotic CHL orbifolds,''
 {{\hypersetup{urlcolor=darkred}\href{https://scipost.org/SciPostPhys.3.1.008/pdf}{SciPost Phys. {\bf 3}, 008 (2017)}\hypersetup{urlcolor=blue}}}  \eprintN{1702.01926}.
  

\bibitem{Fleig:2015vky}
  P.~Fleig, H.~P.~A.~Gustafsson, A.~Kleinschmidt and D.~Persson,
  ``Eisenstein series and automorphic representations,''
  \eprintNM{1511.04265}.

\bibitem{CollingwoodMcGovern}
D.~H. Collingwood and W.~M. McGovern, {\em Nilpotent orbits in semisimple {L}ie
  algebras}.
\newblock Van Nostrand Reinhold Mathematics Series. Van Nostrand Reinhold Co.,
  New York, 1993.

\bibitem{Spaltenstein}
  N.~Spaltenstein,
  {\em Classes unipotentes et sous-groupes de Borel},
  Lecture Notes in Mathematics 946, Springer, 1982.

\bibitem{Borho:1982}
  W.~Borho, J.-L.~Brylinski, 
  ``Differential operators on homogeneous spaces I. Irreducibility of the associated variety for annihilators of induced  modules,''
 \doi{Invent. Math, 69 (1982), 437--476}{doi:10.1007/BF01389364}.


\bibitem{Barbasch:1985}
  D.~Barbasch and D.~A.~Vogan, Jr., 
  ``Unipotent representations of complex semisimple groups,''
  \doi{Ann. of Math. (2), 121 (1985), 41--110}{doi:10.2307/1971193}.

\bibitem{Argurio:1998cp}
  R.~Argurio,
  ``Brane physics in M theory,''
  \eprint{hep-th/9807171}.
  
\bibitem{Witten:2000mf}
  E.~Witten,
  ``BPS Bound states of D0 - D6 and D0 - D8 systems in a B field,''
  \doi{JHEP {\bf 0204} (2002) 012}{doi:10.1088/1126-6708/2002/04/012}
  \eprint{hep-th/0012054}.

\bibitem{Denef:2007vg}
  F.~Denef and G.~W.~Moore,
  ``Split states, entropy enigmas, holes and halos,''
  \doi{JHEP {\bf 1111} (2011) 129}{doi:10.1007/JHEP11(2011)129}
  \eprint{hep-th/0702146}.
  
  
  
\bibitem{Shih:2005qf} 
  D.~Shih, A.~Strominger and X.~Yin,
  ``Counting dyons in $\mathcal{N}=8$ string theory,''
  \doi{JHEP {\bf 0606}, 037 (2006)}{doi:10.1088/1126-6708/2006/06/037}
  \eprint{hep-th/0506151}.

\bibitem{Papageorgakis:2014dma}
  C.~Papageorgakis and A.~B.~Royston,
  ``Revisiting Soliton Contributions to Perturbative Amplitudes,''
  \doi{JHEP {\bf 1409} (2014) 128}{doi:10.1007/JHEP09(2014)128}
  \eprintN{1404.0016}.
  


\bibitem{Jiang:2014}
  D.~Jiang, B.~Liu and G.~Savin,
  ``Raising nilpotent orbits in wave-front sets,''
  \doi{Represent. Theory {\bf 20} (2016) 419--450}{http://dx.doi.org/10.1090/ert/490}
 \eprintNM{1412.8742}.

\bibitem{DHoker:2005vch}
  E.~D'Hoker and D.~H.~Phong,
  ``Two-loop superstrings VI: Non-renormalization theorems and the 4-point function,''
  \doi{Nucl.\ Phys.\ B {\bf 715} (2005) 3}{doi:10.1016/j.nuclphysb.2005.02.043}
  \eprint{hep-th/0501197}.


\bibitem{DHoker:2014oxd}
  E.~D'Hoker, M.~B.~Green, B.~Pioline and R.~Russo,
  ``Matching the $D^{6}R^{4}$ interaction at two-loops,''
  \doi{JHEP {\bf 1501} (2015) 031}{doi:10.1007/JHEP01(2015)031}
  \eprintN{1405.6226}.

\bibitem{Herz}
  C.~S.~Herz,
  ``Bessel functions of matrix argument'',
  \doi{Annals of Math. {\bf 61} (1955) 474--523}{10.2307/1969810}.

\bibitem{Terras}
  A.~Terras,
  ``Harmonic analysis on symmetric spaces and applications II'',
  \doi{Springer Verlag (New York, 1988)}{10.1007/978-1-4612-3820-1}.
  
\bibitem{Kac}
  V.~G.~Kac, {\it Infinite-dimensional Lie algebras}, Cambridge University Press (Third edition, 1995).  
  
  \bibitem{Peterson:1983}
D.~H. Peterson and V.~G. Kac, ``Infinite flag varieties and conjugacy
  theorems,'' {\em Proc. Nat. Acad. Sci. U.S.A.} {\bfseries 80} no.~6 i.,
  (1983) 1778--1782.

\bibitem{Breitenlohner:1986um}
  P.~Breitenlohner and D.~Maison,
  ``On the Geroch Group,''
  Ann.\ Inst.\ H.\ Poincare Phys.\ Theor.\  {\bf 46} (1987) 215.  

  
\bibitem{West:2003fc}
  P.~C.~West,
  ``E(11), SL(32) and central charges,''
  \doi{Phys.\ Lett.\ B {\bf 575} (2003) 333}{doi:10.1016/j.physletb.2003.09.059}
  \eprint{hep-th/0307098}.
  
  \bibitem{Kleinschmidt:2003jf}
  A.~Kleinschmidt and P.~C.~West,
  ``Representations of G+++ and the role of space-time,''
  \doi{JHEP {\bf 0402} (2004) 033}{doi:10.1088/1126-6708/2004/02/033}
  \eprint{hep-th/0312247}.
  
\bibitem{Kumar}
S.~Kumar, \href{http://dx.doi.org/10.1007/978-1-4612-0105-2}{{\em Kac-{M}oody
  groups, their flag varieties and representation theory}}, vol.~204 of {\em
  Progress in Mathematics}.
\newblock Birkh\"auser Boston, Inc., Boston, MA, 2002.

\bibitem{Steinberg}
  R.~Steinberg, \textit{Lectures on Chevalley groups}, University Lecture Series, vol. 66, AMS (Providence, RI, 2016). Notes prepared by John Faulkner and Robert Wilson; Revised and corrected edition of the 1968 original.
  
  
\end{thebibliography}
\end{document}